\pdfoutput=1
\documentclass{vldb}
\usepackage{graphicx}
\usepackage{balance}  
\usepackage{subfigure}
\usepackage{epstopdf}
\usepackage{xcolor}
\usepackage{url}
\usepackage[lined,ruled]{algorithm2e}

\newcommand\yaochang[1]{\textcolor{black}{#1}}

\newcommand{\tabincell}[2]{\begin{tabular}{@{}#1@{}}#2\end{tabular}}
\newtheorem{definition}{Definition}

\begin{document}


\title{DGCC: A New Dependency Graph based Concurrency Control Protocol for Multicore Database Systems}



%
%
%
%

\numberofauthors{1}
\author{Chang Yao$^\ddag$, Divyakant Agrawal$^\sharp$, Pengfei Chang$^\S$, Gang Chen$^\S$\\ Beng Chin Ooi$^\ddag$, Weng-Fai Wong$^\ddag$, Meihui Zhang$^\dag$
\vspace{3mm}\\
\fontsize{10}{10}\upshape
$^\ddag$National University of Singapore, $^\sharp$University of California at Santa Barbara\\
\fontsize{10}{10}\upshape
$^\S$Zhejiang University
\vspace{1.6mm}, $^\dag$Singapore University of Technology and Design\\
\fontsize{10}{10}\upshape
$^\ddag$\{yaochang,ooibc,wongwf\}@comp.nus.edu.sg,
$^\sharp$agrawal@cs.ucsb.edu \\
\fontsize{10}{10}\upshape
$^\S$\{changpeng3336,cg\}@cs.zju.edu.cn, $^\dag$meihui\_zhang@sutd.edu.sg\\
}

\maketitle

\begin{abstract}
Multicore CPUs and large memories are increasingly becoming the norm
in modern computer systems.
However, current database management systems (DBMSs) are generally
ineffective in exploiting the parallelism of such systems.
In particular, contention can lead to a dramatic fall in
performance.
In this paper, we propose a new concurrency control protocol called DGCC (Dependency Graph based Concurrency Control) that separates concurrency
control from execution.
DGCC builds dependency graphs for batched transactions before executing them.
Using these graphs, contentions within the same batch of transactions are resolved before
execution. As a result, the execution of the transactions does not need to deal with
contention while maintaining full equivalence to that of serialized execution.
This better exploits multicore hardware and achieves higher level of parallelism.
To facilitate DGCC, we have also proposed a system architecture that
does not have certain centralized control components yielding better scalability, 
as well as
supports a more efficient recovery mechanism.
Our extensive experimental study shows that DGCC achieves up to four times higher throughput
compared to that of state-of-the-art concurrency control protocols for high contention
workloads.
\end{abstract}

\section{Introduction}

Advancement in multicore processors in the last decade have
enabled programs to significantly improve performance
by exploiting parallelism.
Further, the
availability of larger and cheaper main memory makes it possible for a significant amount of data
to reside in main memory. It is now feasible to
have a single multicore system with large memory to
handle applications that
were previously supported by multiple machines.
However, current database management systems (DBMSs) are not
designed to fully exploit these new hardware features.
In this paper, we will examine the design of
multicore in-memory OLTP systems with the goal of
improving the throughput of transaction processing by better
exploiting modern multicore hardware.
In summary, we divide transactions arriving at the DBMS into batches. 
Every transaction within each batch is chopped up into {\em transaction pieces} which are reorganized
into an efficient concurrent execution plan that has no contention.
We present a new control concurrency protocol based on the dependencies of transactions
that ensures the correctness of the execution.

\begin{figure*}[bt]
\centering
\includegraphics[width=0.95\textwidth]{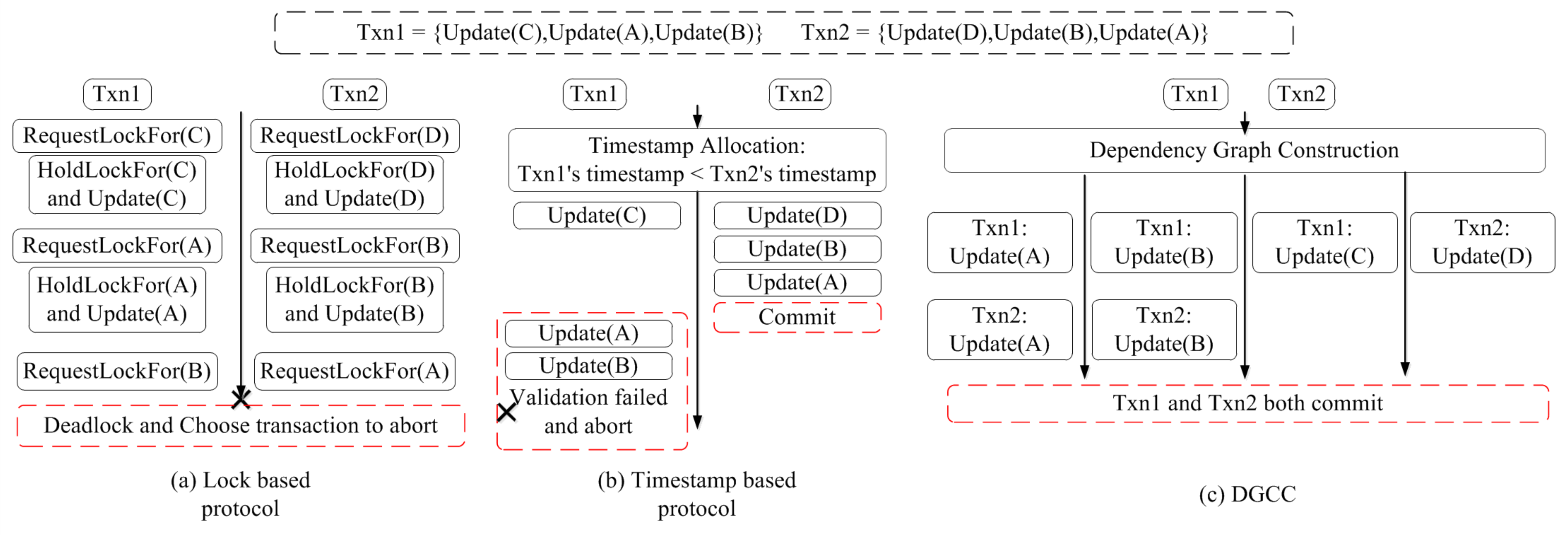}
\caption{An Example with Two Transactions}
\label{fig:example_intro}
\end{figure*}

We call our new concurrency control protocol {\em Dependency Graph based Concurrency Control} (DGCC).
DGCC differs from traditional lock based or timestamp based protocols in
that it separates the logic for concurrency control from the execution of the transactions.
In traditional OLTP systems, each transaction is handled by a worker thread from its beginning to its end.
The worker thread is responsible for contention resolution and execution.
Since each thread consumes systems resources, there is a limit to the
number of threads and hence the number of concurrent transactions that can be present at
any one time. Furthermore, overall performance is affected by contention as well
as the inability to fully exploiting parallelism.
To alleviate the problem and improve scalability,
DGCC first chops up a batch of transactions into transaction pieces, and then builds a {\em dependency graph}
that incorporates the dependency relationship of the transaction operations.
DGCC then executes these dependency graphs in a manner that
guarantees the execution of the operations is serializable. 
Furthermore, the execution will have no contention at runtime.

We illustrate the basic idea of DGCC and compare it with the two traditional concurrency control protocols in Figure~\ref{fig:example_intro}.
For a lock based protocol, as shown in Figure~\ref{fig:example_intro}(a),
a deadlock occurs when transaction $Txn1$ is holding {\em A}'s lock and requesting {\em B}'s lock, while
transaction $Txn2$ is holding {\em B}'s lock and requesting {\em A}'s lock.
To break the deadlock, either
transaction $Txn1$ or transaction $Txn2$ must be aborted.
In a timestamp based protocol, shown in Figure~\ref{fig:example_intro}(b),
transaction $Txn1$'s operations overlap
with transaction $Txn2$'s operations. At the validation phase of
transaction $Txn1$, it is found that
record $A$ has been modified by transaction $Txn2$, which completed after transaction $Txn1$
started and had committed earlier. This causes transaction $Txn1$ to be aborted.
In addition, in both lock based and timestamp based protocols,
operations in one transaction must run sequentially within a single thread.
As such, the two transactions in Figure~\ref{fig:example_intro} (a) and (b) can be
concurrently executed by at most two threads.
In DGCC, during dependency graph construction phase, transactions are broken down into transaction pieces,
which allows the system to parallelize the execution at level of operations.
More specifically,
DGCC enables concurrent execution of the transaction operations as long as the they do not conflict.
As shown in Figure~\ref{fig:example_intro} (c),
four threads are initiated for transaction $Txn1$ and transaction $Txn2$'s execution as they can simultaneously operate on four different records.
If there are operations with dependency (e.g., read and write records $A$ and $B$ from the two transactions),
DGCC will execute them in order.
Finally, both transactions will successfully commit.
In this manner, DGCC reduces the abort rate while at the same time
enabling higher concurrency and guaranteeing serializability.

DGCC consists of a graph construction phase and an execution phase,
using a different work partitioning strategy for each phase.
In particular,
one worker thread is responsible for the construction of each dependency graph.
At graph construction phase, $n$ worker threads will work in parallel to build $n$
different dependency graphs at the same time.
If more than one transaction attempt to access the same data, during the execution phase,
the dependency graphs constructed by DGCC guarantee that they will be executed in a serialized manner.
In general, however, this approach exposes parallelism when the opportunity presents itself.

DGCC is based on batch processing in a multicore in-memory system.
As with any batch processing, latency is a valid concern.
However, we shall reason that this is feasible in practice.
First, in real applications,
requests at the client side are always sent to the server in batches
so as to reduce the network overhead. 
More importantly,
in-memory systems always need to write transaction logs to disk for the
purpose of reliability.
In order to reduce disk I/O cost,
group commit protocols \cite{dewitt1984implementation} are not uncommon. 
In other words, current systems already both receive and commit transaction in a batch manner.
Secondly, in the context of in-memory multicore systems,
data access is extremely fast compared to that in traditional disk-based systems,
thereby reducing latency.
Thirdly,
the latency due to the batch processing can actually be minimized by the
tuning of the batch size. 
In summary, if the execution strategy is well designed, latency can be
controlled to within acceptable bounds.
The experiments conducted in our performance study confirms that
fast batch processing is achievable.

We have implemented an in-memory OLTP system with DGCC concurrency control
protocol that
supports high concurrency, efficient recovery and good scalability.
The system architecture is designed for the modern multicore environment.
Our experiments show that it
achieves significantly higher throughput, and scales well
compared to other concurrency control protocols.

In summary, this paper makes the following contributions:
\begin{itemize}
\item We propose DGCC, a new concurrency control protocol that separates contention resolution from execution using dependency graph and achieves higher parallelism.

\item
A new in-memory multicore OLTP system supporting DGCC is prototyped.
Besides DGCC, it supports an efficient recovery mechanism and
a customized memory allocation scheme that helps to avoid system memory malloc
at the runtime.

\yaochang{
}

\item An extensive performance study of DGCC against three state-of-the-art concurrency control protocols was conducted.
The performance study using two benchmarks shows that DGCC achieves up to four times higher throughput than the other three concurrency control protocols.

\end{itemize}

The remainder of the paper is organized as follows. In Section 2, we introduce
classical concurrency control protocols. We present DGCC in Section 3, and the architecture of
our prototype system in Section 4. A comprehensive evaluation is presented in Section 5,
and we review some related
work in Section 6. Finally, the paper is concluded in Section 7.

\section{Existing Concurrency Control Protocols}
A transaction in a DBMS consists of a sequence of read and write operations.
The DBMS must guarantee that
(a) only serializable and recoverable schedules are allowed, (b) no operations of committed transactions are lost,
and (c) the effects of partial transactions are not retained.
In short, the DBMS is responsible to ensure the ACID (Atomicity, Consistency, Isolation and Durability) \cite{haerder1983principles} properties.

In the multicore era,
concurrency control protocols should enable multi-user programs to be interleaved and executed concurrently with
the net effect being is identical to executing them in a serial order.
Essentially, concurrency control protocols ensure the atomicity and isolation properties.
Many research efforts have been devoted to this area.
We shall follow the canonical categorization in ~\cite{yustaring} and
review them in two categories, namely lock and timestamp based protocols.

\subsection{Lock Based Protocols}
The essential idea of lock based protocols is making use of locks
to control the access to data.
A transaction must acquire a lock on an object before it
can operate on the object to prevent unsafe interleaving of transactions.
With this kind of protocols,
transactions accessing data locked by other transactions may be blocked
until the requested locks are released.
There are at least two types of locks: write lock and read lock.
Write lock is an exclusive lock and read lock can be a shared lock.
The rules of lock blocking is usually presented by lock compatibility
table~\cite{ramakrishnan2003database}.

System with lock based protocol may use a global lock manager to
grant and release locks. To improve the scalability,
de-centralized lock manager has been proposed that co-locate the
lock table with the raw data.


Two-phase locking(2PL)~\cite{bernstein1979formal,eswaran1976notions} is a widely used
locking protocol.
In the growing phase, a transaction first acquires locks
without releasing any. During the
shrinking phase, it can only release
locks without acquiring any locks.
In a multi-programmed environment,
lock based protocols have to deal with deadlocks, and transactions may be aborted when a deadlock cannot be prevented.
Overall system performance is affected by transaction blocking,
deadlock detection and resolving.

\subsection{Timestamp Based Protocol}
Timestamp based protocols~\cite{bernstein1981concurrency,bernstein1987concurrency}
assigns a global timestamp before processing.
By ordering the
timestamp, the execution order of transactions is determined.
When multiple transactions attempt to access the same data,
the transaction with smaller timestamp should be executed first.
As shown in Figure \ref{fig:example_intro}, if conflicts exist
during execution, the transaction will be aborted and restarted.

Optimistic Concurrency Control (OCC)~\cite{kung1981optimistic} and Multi-Version Concurrency Control
(MVCC)~\cite{bernstein1983multiversion}
are two widely used timestamp based protocols.
OCC assumes low data contention where conflicts are rare.
Transactions can complete without blocking.
However,
before a transaction commits, a validation is performed to check if there is
any conflict. If conflicts exist, the transaction will be aborted and restarted.
MVCC maintains multiple versions of each data object and is more efficient for read operations.
The read operations can access the data of an appropriate version without being blocked by other write operations.
A periodic garbage collection is required to free inactive data.

Timestamp based protocols perform poorly on workloads with high contention,
due to their high abort rate.
Aborts not only consume computing resources, but also
additional work needs to be performed to undo the aborted transactions.
Moreover, these kinds of protocols usually requires a centralized manager
to assign unique timestamp to transactions.
This limits system scalability.

\section{Dependency Graph based Concurrency Control}
In this section, we present the Dependency Graph based Concurrency Control
(DGCC) protocol.

Typically, arriving transactions cannot be processed by the system immediately.
They will first wait in a transaction queue.
Unlike the worker thread in the lock and timestamp based concurrency control protocols which
processes the transactions one by one,
DGCC grabs a batch of transactions from the transaction queue to process.
The batch size depends on the number of transactions in the transaction queue and
the pre-defined maximal batch size.
There are two separate phases: \textbf{Dependency Graph Construction} and
\textbf{Dependency Graph Execution}.
Multi-threading is used in both phases for maximal parallelism.
More importantly, no locks are required in the whole process.
Neither are there any aborts due to conflicts.
\begin{table}[tb]
\caption{Notations}
\begin{tabular}{|l|l|} \hline
$\mathcal{T}$ & a set of transaction\\ \hline
$\mathcal{G}$ & dependency graph\\ \hline
$s$ & a schedule of transaction execution\\ \hline
$G(s)$ & conflict graph of s\\ \hline
$t_i$ & a transaction with time stamp $i$\\ \hline
$\Phi_{t_i}$ & the set of pieces of transaction $t_i$\\ \hline
$\phi^p_{t_i}$  & the $p_{th}$ piece of transaction $t_i$\\ \hline
\emph{readset}($\phi^p_{t_i}$) & the read record set of $\phi^p_{t_i}$\\ \hline
\emph{writeset}($\phi^p_{t_i}$) & the write record set of $\phi^p_{t_i}$\\ \hline
\emph{accessset}($\phi^p_{t_i}$) & \emph{readset}($\phi^p_{t_i}$) $\cup$ \emph{writeset}($\phi^p_{t_i}$)\\ \hline
$k$ & one record stored in database\\ \hline
$\mathcal{L}(k)$ & latest write transaction piece on $k$\\ \hline
$\Psi(k)$ & the dominating set of $k$\\ \hline
$\succ_{\mbox{\scriptsize time-order}}$ & timestamp ordering dependency\\ \hline
$\succ_{\mbox{\scriptsize logic}}$ & logic dependency\\ \hline
\end{tabular}
\label{tab:table1}
\end{table}
Table~\ref{tab:table1} summarizes the notations used in this section

\subsection{Chopping Transactions in DGCC}
Conventional concurrency control protocols process a single transaction sequentially
with no concurrent processing within a transaction.
DGCC chops a transaction into a set of smaller {\em transaction pieces} according to its type and internal logics.
Transactions in OLTP applications are often repetitive and {\em store-procedures} are widely
used in current systems.
A transaction piece consists of a set of store-procedures that operates on
some records in the database.
Each piece is represented as a vertex in our dependency graph.
It is the unit in both the dependency graph construction
and dependency graph execution.
Transaction pieces may be partially ordered.
We define the partial-order between two transaction pieces as a logic dependency
in the following subsection.

\subsection{Dependency Graph Construction}

\begin{figure*}[hbt]
\centering
\includegraphics[width=0.9\linewidth]{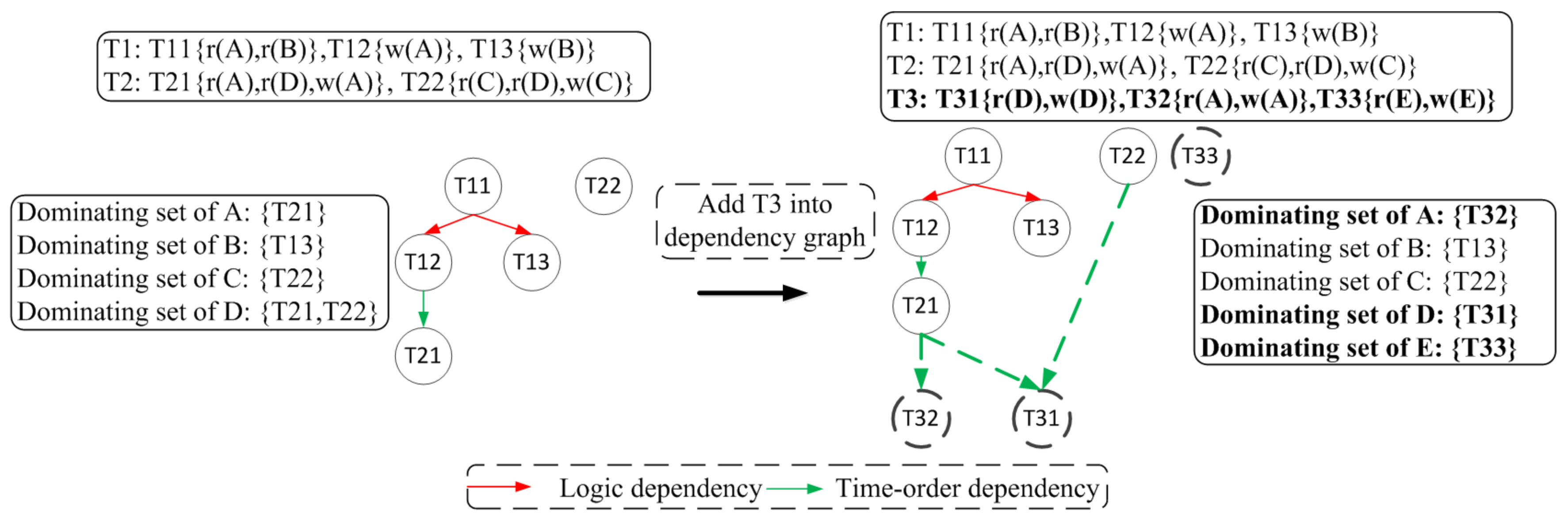}
\caption{Dependency Graph Construction}
\label{fig:construct}
\end{figure*}

During dependency graph construction,
one batch of transactions is divided into several disjoint sets of transactions.
A worker thread will construct
a dependency graph $\mathcal{G}$ from a set
of transactions $\mathcal{T} = \{t_1,t_2,\cdots,t_n\}$.
Each transaction $t_i$ is associated with a timestamp $i$.
Transactions in a given set are processed ordered by their timestamps.
Each transaction, $t_i$, is further divided into a set of transaction pieces.
$\Phi_{t_i}$ = \{${\phi^1_{t_i}},{\phi^2_{t_i}},\cdots,{\phi^m_{t_i}}$\}.
We define two types of dependency relations on the pieces: {\em logic dependency relation} $\succ_{\mbox{\scriptsize logic}}$ and
{\em timestamp ordering dependency} relation $\succ_{\mbox{\scriptsize time-order}}$.
We first define the logic dependency relation $\succ_{\mbox{\scriptsize logic}}$:

\begin{definition}[Logic Dependency]
Transaction piece ${\phi^q_{t_j}}$ logically depends on ${\phi^p_{t_i}}$, denoted as
${\phi^q_{t_j}} \succ_{\mbox{\scriptsize logic}} {\phi^p_{t_i}}$, if and only if $i=j$ and
${\phi^q_{t_j}}$ is executed {\em after} ${\phi^p_{t_i}}$.
\end{definition}

From the above definition, we can see that $\succ_{\mbox{\scriptsize logic}}$ represents the logical execution
order of the pieces within one transaction.
Apart from the logic dependency relation, we also need to resolve the execution order of pieces from different transactions,
which is defined by timestamp ordering dependency relation $\succ_{\mbox{\scriptsize time-order}}$.
For a transaction piece $\phi^p_{t_i}$,
$\emph{writeset}({\phi^p_{t_i}})$ and $\emph{readset}({\phi^p_{t_i}})$ are used to represent the set of
records written to and read, respectively.
The access set
$\emph{accessset}({\phi^p_{t_i}})$ is \emph{readset}($\phi^p_{t_i}$) $\cup$ \emph{writeset}($\phi^p_{t_i}$).

\begin{definition}[Timestamp Ordering Dependency]
A timestamp ordering dependency ${\phi^q_{t_j}} \succ_{\mbox{\scriptsize time-order}} {\phi^p_{t_i}}$ exists
if and only if $j>i$ and ($\mbox{writeset}({\phi^q_{t_j}}) \cap \mbox{accessset}({\phi^p_{t_i}}) \neq \O$ or
$\mbox{accessset}({\phi^q_{t_j}}) \cap \mbox{writeset}({\phi^p_{t_i}}) \neq \O$).
\end{definition}

\begin{definition}[Dependency Graph]
Given a set of transactions $\mathcal{T} = \{t_1,t_2,\cdots,t_n\}$, and the associated
sets of transaction pieces $\Phi_{t_1},{\Phi_{t_2}},\cdots,\Phi_{t_n}$,
the dependency graph $\mathcal{G} = (\mathcal{V},\mathcal{E})$ consists of
\begin{itemize}
\item {$\mathcal{V} = \Phi_{t_1} \cup {\Phi_{t_2}} \cup \cdots \cup\Phi_{t_n}$}, and
\item {$\mathcal{E} = \{(\phi^p_{t_i},{\phi^q_{t_j}})$ such that
$\phi^p_{t_i} \in \Phi_{t_i}$,
$\phi^q_{t_j} \in \Phi_{t_j}$, and
$\phi^q_{t_j} \succ_{\mbox{\scriptsize logic}}\phi^p_{t_i}$ or
$ \phi^q_{t_j} \succ_{\mbox{\scriptsize time-order}} \phi^p_{t_i}$\}}.
\end{itemize}
\end{definition}

It is not efficient to analyze ${\phi^q_{t_j}}$ with every piece in $\mathcal{G}$ when we add ${\phi^q_{t_j}}$ into $\mathcal{G}$.
Furthermore, explicitly recording all timestamp ordering dependency edges between all the transaction pieces will result in a lot of edges.
So during dependency graph construction, we maintain the dominating set $\Psi(k)$ for each record $k$ that is accessed in $\mathcal{G}$.
Here we define the latest write transaction piece on $k$ as:
\begin{definition} [Latest Write Transaction Piece]
$\mathcal{L}(k) = \phi^p_{t_i} such\ that\ \nexists \phi^q_{t_j}\in\mathcal{V},(j > i)\ and\ k \in \mbox{writeset}(\phi^q_{t_j})$
\end{definition}

Then the dominating set $\Psi(k)$ is defined as follows:
\begin{definition}[Dominating Set]
$\Psi(k) =
\{\phi^p_{t_{i}} |\phi^p_{t_{i}} = \mathcal{L}(k)\ and\ (\nexists {\phi^q_{t_{j}}} \in \mathcal{V}, (j>i) \wedge k \in accessset({\phi^q_{t_{j}}}))\}$ $\cup$
$\{{\phi^p_{t_{i}}}|k \in readset({\phi^p_{t_{i}}})\ and\ (\nexists {\phi^q_{t_{j}}} \in \mathcal{V}, (j \geq i) \wedge k \in writeset({\phi^q_{t_{i}}}))\}$
\end{definition}

The dominating set $\Psi(k)$ contains only $\mathcal{L}(k)$ when there are no subsequent pieces accessing $k$ or it will
contain all the operations that read $k$ after $\mathcal{L}(k)$.
Hence by maintaining the dominating set $\Psi(k)$ for each record $k$,
we only need to analyse ${\phi^q_{t_j}}$ with the transaction pieces in $\Psi(k)$ to add edges
when we insert ${\phi^q_{t_j}}$ into $\mathcal{G}$.

Now, we can summarize the dependency graph construction algorithm for a set of transactions $\mathcal{T}$ as Algorithm \ref{algo:dgccconstruct}.
We use the example in Figure \ref{fig:construct} to illustrate the dependency graph construction process.
There are three transactions $T1$, $T2$, $T3$.
Our example begins after $T1$ and $T2$ have already been inserted into the dependency graph.
The red directed edges represent logical dependency and green directed edges represent timestamp ordering dependency.

When $T3$ is inserted into the dependency graph,
it is divided into three pieces $T31$, $T32$ and $T33$.
For $T31$, we check the dominating set of record D add green directed edges from $T21$ to $T31$ and from $T22$ to $T31$.
For $T32$, we check the dominating set of record $A$ and add green directed edges from $T21$ to $T32$.
For $T33$, there is no dominating set of record $E$ and hence we just insert $T33$ into $\mathcal{G}$
with no edges connected to it.
Apart from adding edges into $\mathcal{G}$,
we update the dominating set according to the $\emph{accessset}$ of each piece.

\begin{algorithm}[h]
 \caption{construct the dependency graph $\mathcal{G}$ for one transaction set $\mathcal{T}$}\label{algo:dgccconstruct}
 \BlankLine
 \For{$t_j$ in $\mathcal{T}$}{
 split $t_j$ as $\Phi_{t_j}$ = \{$\phi^1_{t_j},\phi^2_{t_j},\cdots,\phi^m_{t_j}$\}\;
    \For{$\phi^q_{t_j}$ in $\Phi_{t_j}$}{
    \For{$k_h$ in \emph{accessset}($\phi^q_{t_j}$)}{
    \If{$\Psi(k_h)$ = $\O$}{
        add $\phi^q_{t_j}$ into $\mathcal{G}$ and insert $\phi^q_{t_j}$ into $\Psi(k_h)$\;
        break\;
    }
    \If{$\Psi(k_h)$ contains only one piece $\phi^p_{t_i}$ that write on $k_h$}{
    add edge from $\phi^p_{t_i}$ point to $\phi^q_{t_j}$ representing $\phi^q_{t_j} \succ_{\mbox{\scriptsize time-order}} \phi^p_{t_i}$\;
    clear $\Psi(k_h)$ and insert $\phi^q_{t_j}$ into $\Psi(k_h)$\;
    }\Else{
     \If{$\phi^q_{t_j}$ read on $k_h$}{
     add edge from $\mathcal{L}(k_h)$ point to $\phi^q_{t_j}$ representing $\phi^q_{t_j} \succ_{\mbox{\scriptsize time-order}}\mathcal{L}(k_h)$\;
     insert $\phi^q_{t_j}$ into $\Psi(k_h)$;
     }\Else{
       \For{$\phi^p_{t_i}$ in $\Psi(k_h)$}{
       add edge from $\phi^p_{t_i}$ point to $\phi^q_{t_j}$ representing $\phi^q_{t_j} \succ_{\mbox{\scriptsize time-order}} \phi^p_{t_i}$\;
       }
       clear $\Psi(k_h)$ and insert $\phi^q_{t_j}$ into $\Psi(k_h)$\;
     }
     }
    }
    }
    add edges based on $\succ_{\mbox{\scriptsize logic}}$ dependency\;
 }
\end{algorithm}

\begin{figure}[bt]
\centering
\includegraphics[width=0.8\linewidth]{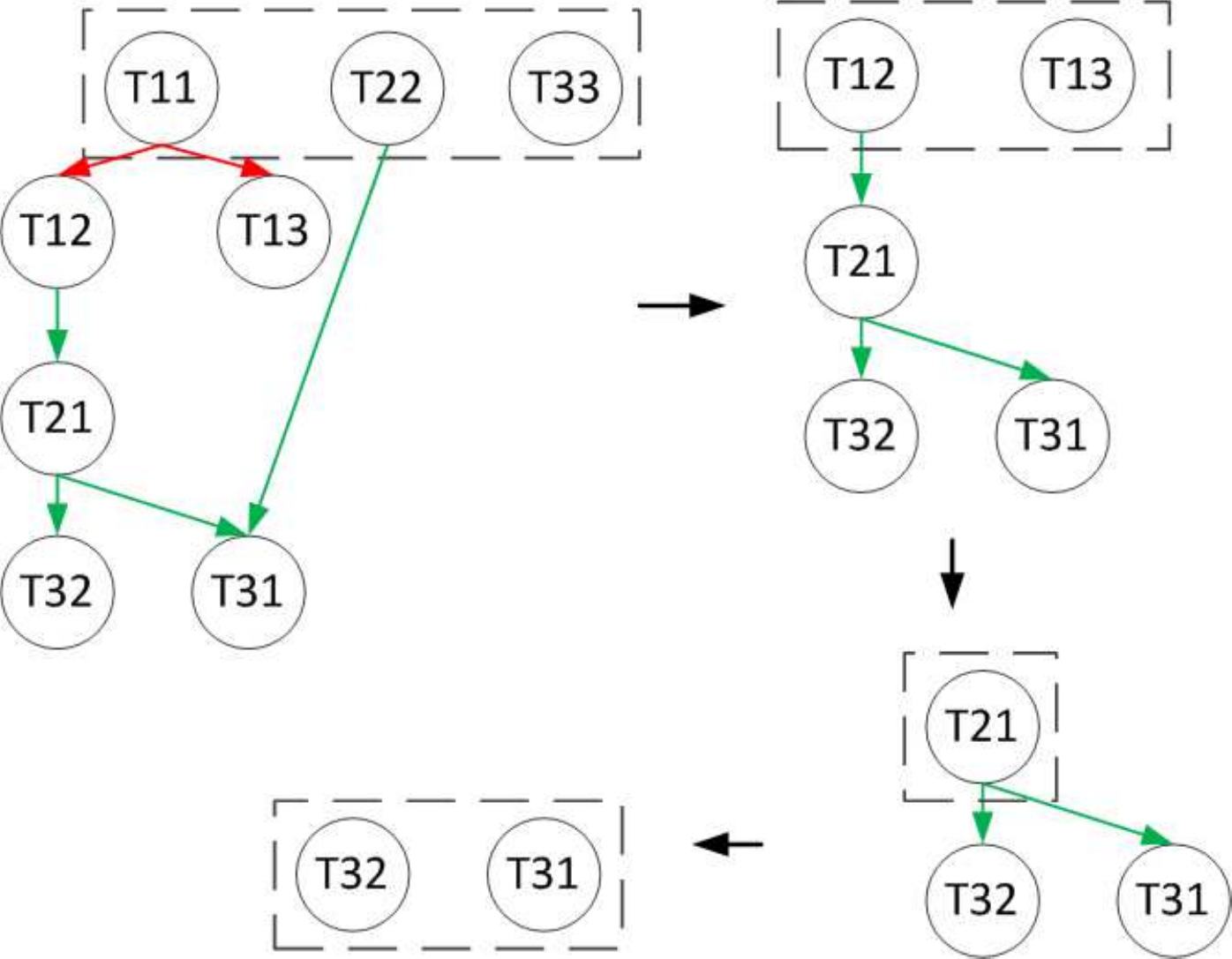}
\caption{Dependency Graph Execution}
\label{fig:execute}
\end{figure}
\subsection{Dependency Graph Execution}
DGCC executes dependency graphs sequentially in a greedy manner.
For a dependency graph $\mathcal{G}$,
we iteratively select vertices with zero in-degree to execute and remove these vertices as well as their
out-going edges from the graph.
This process will repeat until there are no vertices left in $\mathcal{G}$.
We outline the dependency graph execution in
Algorithm \ref{algo:dgccexecution}.
As Figure \ref{fig:execute} shows, at the first round,
we choose $T11$,$T22$, and $T33$ to execute and remove their out-going edges. We then iteratively
select $\{T12,T13\}$,$\{T21\}$,$\{T32,T31\}$ to execute.

\begin{algorithm}[h]
 \caption{execute one dependency graph $\mathcal{G}$}\label{algo:dgccexecution}
 \BlankLine
 \While{\emph{true}}{
 select vertices with zero in-degree as \{$v_1,v_2,\cdots,v_n$\}\;
 \For{$v_i$ in \{$v_1,v_2,\cdots,v_n$\}}{
 add $v_i$'s corresponding piece $\phi^p_{t_i}$ into thread pool\;
 }
 wait for thread pool have no more pieces to execute\;
 }
\end{algorithm}

\subsection{Correctness}
We shall now prove that DGCC guarantees strict serializability.

\noindent
\subsubsection{\textbf{Conflict Serializability}}

In the previous section, the dependency graph $\mathcal{G}$ works as a
schedule $s$ of $\mathcal{T}$.
We can prove that the schedule, $s$, is conflict-serializable based on
{\em Conflict Serializability Theorem}\cite{weikum2001transactional}.
In other words, we need to show that its {\em conflict graph} $G$($s$) is acyclic.

\begin{definition}[Conflict Graph]
Let $s$ be a schedule. The Conflict Graph, $G$($s$) = ($V$,$E$) of $s$, is defined by
\begin{displaymath} V = \mathcal{T}  \end{displaymath}
\begin{displaymath} (t_i,t_j) \in E \iff (i \neq j)\ and\ \exists \phi^p_{t_i}, \phi^q_{t_j}\in \mathcal{V},
\phi^q_{t_j} \succ \phi^p_{t_i}
\end{displaymath}
\end{definition}

As we only have two dependency relations $\succ_{\mbox{\scriptsize time-order}}$ and $\succ_{\mbox{\scriptsize logic}}$,
the conflict relation in the conflict graph $G(s)$ should be either $\succ_{\mbox{\scriptsize time-order}}$ or
$\succ_{\mbox{\scriptsize logic}}$.

Firstly, let's consider $\succ_{\mbox{\scriptsize time-order}}$ in $G(s)$.
Based on its definition,
if there is a directed edge from
$\phi^p_{t_i}$ to  $\phi^q_{t_j}$.
then the timestamp of the second piece, $j$, must be greater than that of the first piece, i.e., $i$.
Now if $G$($s$) is {\em cyclic}, then we can always find a cycle with edges that
$(\phi^{p_0}_{t_{i_0}}, \phi^{p_1}_{t_{i_1}})$,
$(\phi^{p_1}_{t_{i_1}}, \phi^{p_2}_{t_{i_2}})$, $\cdots$,
$(\phi^{p_{v-1}}_{t_{i_{v-1}}}, \phi^{p_v}_{t_{i_v}})$,$(\phi^{p_{v-1}}_{t_{i_{v}}}, \phi^{p_v}_{t_{i_0}})$
where $i_0 < i_1 < \cdots < i_{v-1} < i_{v}$ and $i_v < i_{0}$.
Obviously, this violates the initial condition, namely $i < j$.
In other words, if we only consider the $\succ_{\mbox{\scriptsize time-order}}$ dependency, $G$($s$) must
be acyclic.

Next, we consider $\succ_{\mbox{\scriptsize logic}}$ dependency.
Based on its definition, $\succ_{\mbox{\scriptsize logic}}$
will not lead to an edge in $G$($s$) because $\succ_{\mbox{\scriptsize logic}}$
only exists between two pieces within the same transaction.
So $G$($s$) is still acyclic.

Thus having considered the only two possible forms of dependencies, we can
conclude that
$G$($s$) must be acyclic and $s$ is a conflict-serializable schedule.

\noindent
\subsubsection{\textbf{Strictness}}

In a dependency graph construction,
we have resolved all the conflicts between transactions.
Therefore in executing a dependency graph,
there would not be any transaction abort caused by conflicts.
Transactions can only be aborted due to updates violating the database's schema constraints.
For these, we add condition-variable-check transaction pieces.
As an optimization, if there is more than one condition-variable-check transaction piece,
we will combine them together.
\yaochang{
$\succ_{\mbox{\scriptsize logic}}$ dependency relations are inserted between 
the other pieces in the transaction with the condition-variable-check piece.
If the condition-variable-check piece aborts,
no other pieces in the same transaction that have $\succ_{\mbox{\scriptsize logic}}$ dependency relations with it will execute.}
As a consequence,
no cascading aborts are possible during the execution of a dependency graph.

\subsection{Differences With Transaction Chopping}
{\em Transaction chopping}~\cite{shasha1995transaction} is a method that divides transactions into pieces to
execute with the aim of achieving better parallelism.
It guarantees the serializability of transaction execution by performing static analysis on the
relations between transaction pieces.
This is known as SC-graph analysis.
However a simple static chopping of transactions usually leads to multiple SC-cycles
that have to be merged.
Hence transaction pieces are still relatively large.
DGCC analyzes the relationships between
transaction pieces during runtime,
yielding smaller transaction pieces.
This finer granularity in DGCC, in general, yields more parallelism
than transaction chopping.
Furthermore, during the execution of the transaction pieces,
transaction chopping still requires traditional concurrency control
to resolve conflicts. This leads to possible abort and restart of transaction pieces.
In DGCC's dependency graph execution, no transaction pieces will abort due to conflicts.

\begin{figure}[bht]
\centering
  \subfigure[Transaction Chopping by SC-graph]{
    \label{fig:chop:sc_graph} 
 	  \includegraphics[width=0.46\textwidth,height=0.18\textwidth]{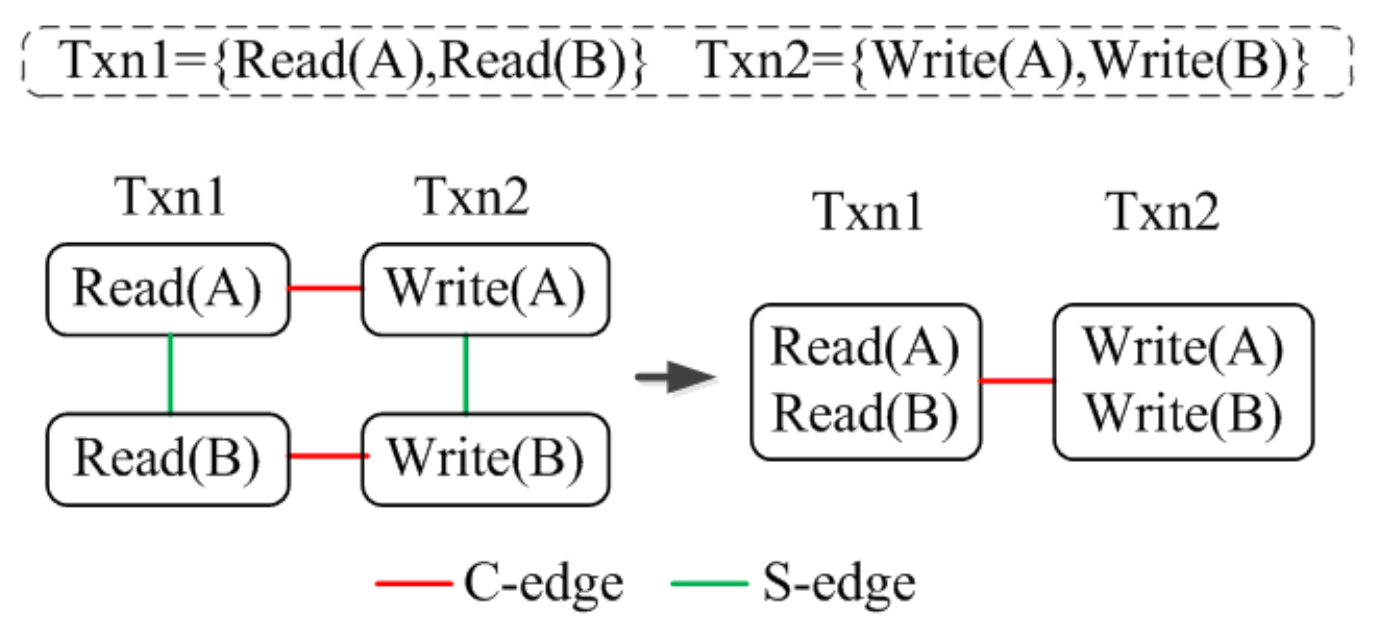}}
  \vspace{.1in}
  \subfigure[Transaction Chopping in DGCC]{
    \label{fig:chop:dgcc} 
    \includegraphics[width=0.4\textwidth,height=0.12\textwidth]{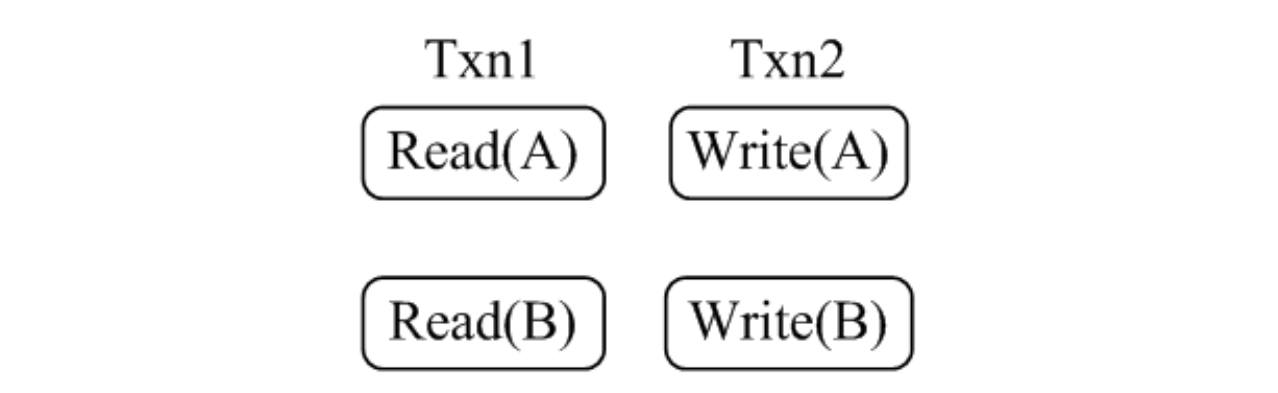}}
  \caption{Transaction Chopping}
  \label{fig:chop}
\end{figure}

Two transactions are shown in Figure \ref{fig:chop}, where
transaction $Txn1$ reads record $A$ and record $B$ while transaction $Txn2$ writes record $A$ and record $B$.
Figure \ref{fig:chop:sc_graph} shows how transaction chopping works with a SC-graph.
SC-cycles in SC-graph should be merged.
Finally, there is only one piece for transaction $Txn1$ and one for transaction $Txn2$.
On the contrary, as illustrated in Figure \ref{fig:chop:dgcc},
\yaochang{DGCC can chop both $Txn1$ and $Txn2$ into two pieces}, which means
fine-grained chopping is acceptable in DGCC.

\section{System Architecture}
\begin{figure}
\centering
\includegraphics[width=0.5\textwidth,height=0.48\textwidth]{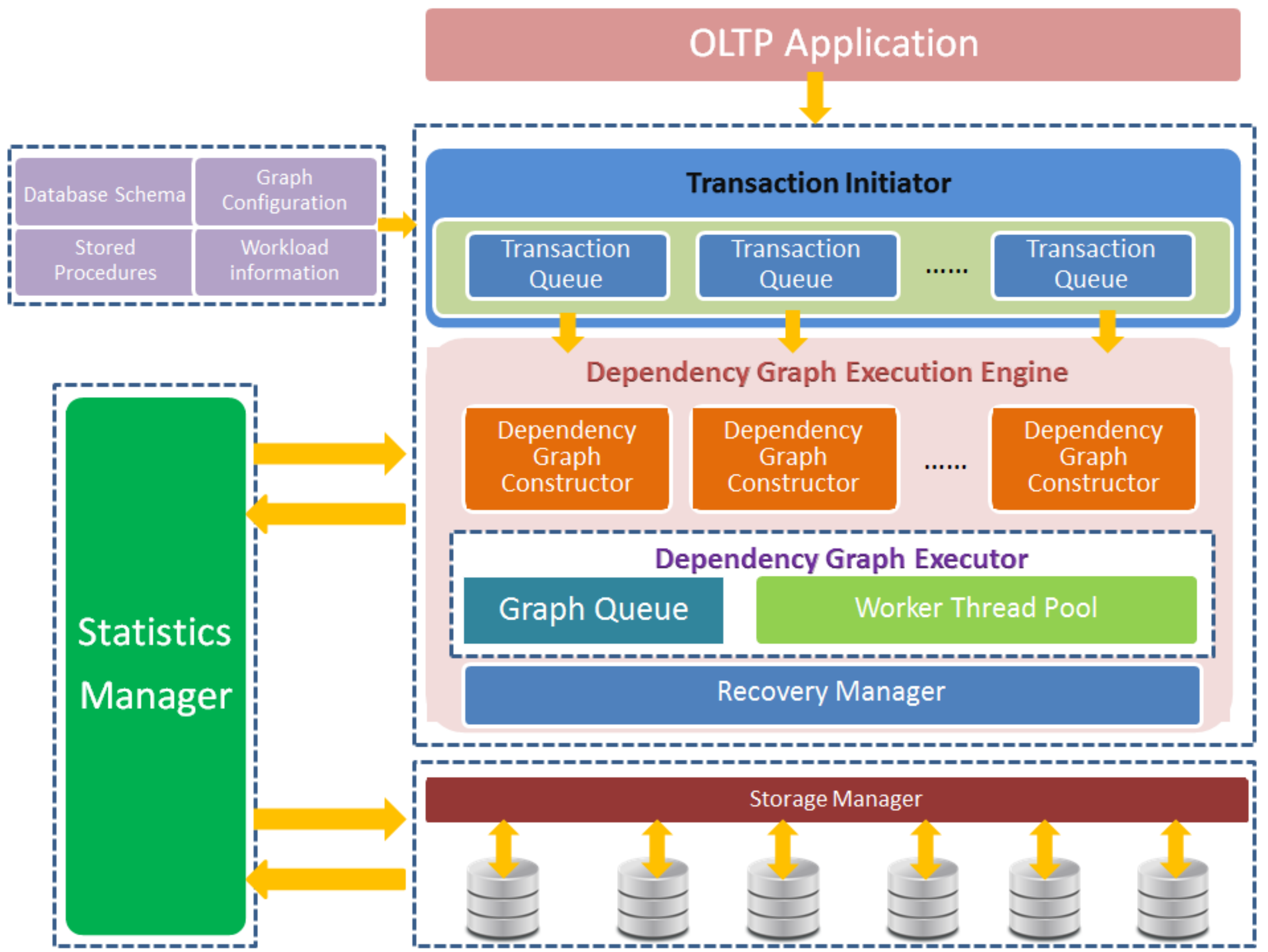}
\caption{System Architecture}
\label{fig:architecture}
\end{figure}

This section presents the architecture of the transaction processing system we have designed to support DGCC.
The system architecture consists of three major components (shown in Figure~\ref{fig:architecture}),
namely Execution Engine, Storage Manager, and Statistic Manager.

\subsection{Execution Engine}

\subsubsection{Initiator}

The execution engine is mainly responsible for managing transaction requests.
It maintains a set of request queues,
and each queue is handled by a dependency graph constructor.
In some applications, transaction requests may have different priorities.
The initiator will adjust the priority of each queue according to requirement, e.g.,
requests of higher priority will be inserted into the queue with higher priority.
At the execution time, requests in the queue with a higher priority will be processed first.
By default, a transaction's priority is set according to its timestamp, i.e., a transaction with a
smaller timestamp has a higher priority.

\subsubsection{Dependency Graph Constructor}
The constructor takes a batch
of transactions from a queue and resolves their contentions by building a dependency graph.
\yaochang{
The batch size is the smaller of the number of transactions in the transaction queue and a pre-defined maximum batch size.
When system is saturated, the batch size is equal to the maximum batch size.
However, we cannot assume that the system is always saturated.
After finishing one round of batch processing,
the constructor will check the transaction queue.
If the number of transactions waiting in the transaction queue is less than
the pre-defined maximum batch size,
all the available transactions will be processed in this batch.
The batch size in our system can be adjusted dynamically to suit workloads
of different request rates.
This strategy ensures that the system will not wait 
indefinitely for sufficient number of transactions to arrive before processing them.
}

For each transaction in the batch,
it first generates vertices according to the transaction's type and its parameters.
To avoid any contention, the dependency graph constructor uses one
single thread to build each dependency graph.
To better exploit parallelism in the CPU, several graphs can be
constructed in parallel by different threads. Each graph construction
is completely independent thereby eliminating any need for synchronization between the
different threads.
\yaochang{
It is possible that there are still conflicts between the different dependency graphs.
We resolve this kind of conflicts by processing the constructed dependency
graphs sequentially at a time in the Graph Executor.}

\subsubsection{Graph Executor}

After graph construction,
the graph executor will execute the graphs according to their priorities.
From the dependency graph,
the executor iteratively extracts an {\em executable vertex set}
consisting of vertices with no incoming edges.
The update of these vertices does not depend on any other vertices.
It follows that any two vertices in the executable vertex set have no contention.
It is therefore safe to allow multiple worker threads to execute the vertices in
the executable vertex set, and they can do so
without requiring any coordinations.
When all the vertices of one graph are processed,
the transactions will commit and responses will be
sent to their clients.

In our prototype, we implemented a fixed number of threads that
will compete to work on either the graph construction or execution.
\yaochang{
During dependency graph execution, if the executable vertex set at each iteration is relatively
small, 
the overhead of context switching and competition among the worker threads
compared to the small amount of work will make multithreading unprofitable.
As an optimization,
if the size of the executable vertex set is small,
we assign all the work to one worker thread
instead of allowing all the worker threads to compete.
}

%
%

\subsection{Recovery Manager}
By maintaining all data in main memory,
in-memory systems significantly reduce disk I/Os,
and, consequently, achieves better throughput with lower latencies.
However, for reliability,
most in-memory systems flush transaction logs to disks and
perform checkpointing periodically.

\begin{figure*}[ht]
\centering
  \subfigure[2PL]{
    \label{fig:wr:2pl} 
    \includegraphics[width=0.235\linewidth,height=0.22\linewidth]{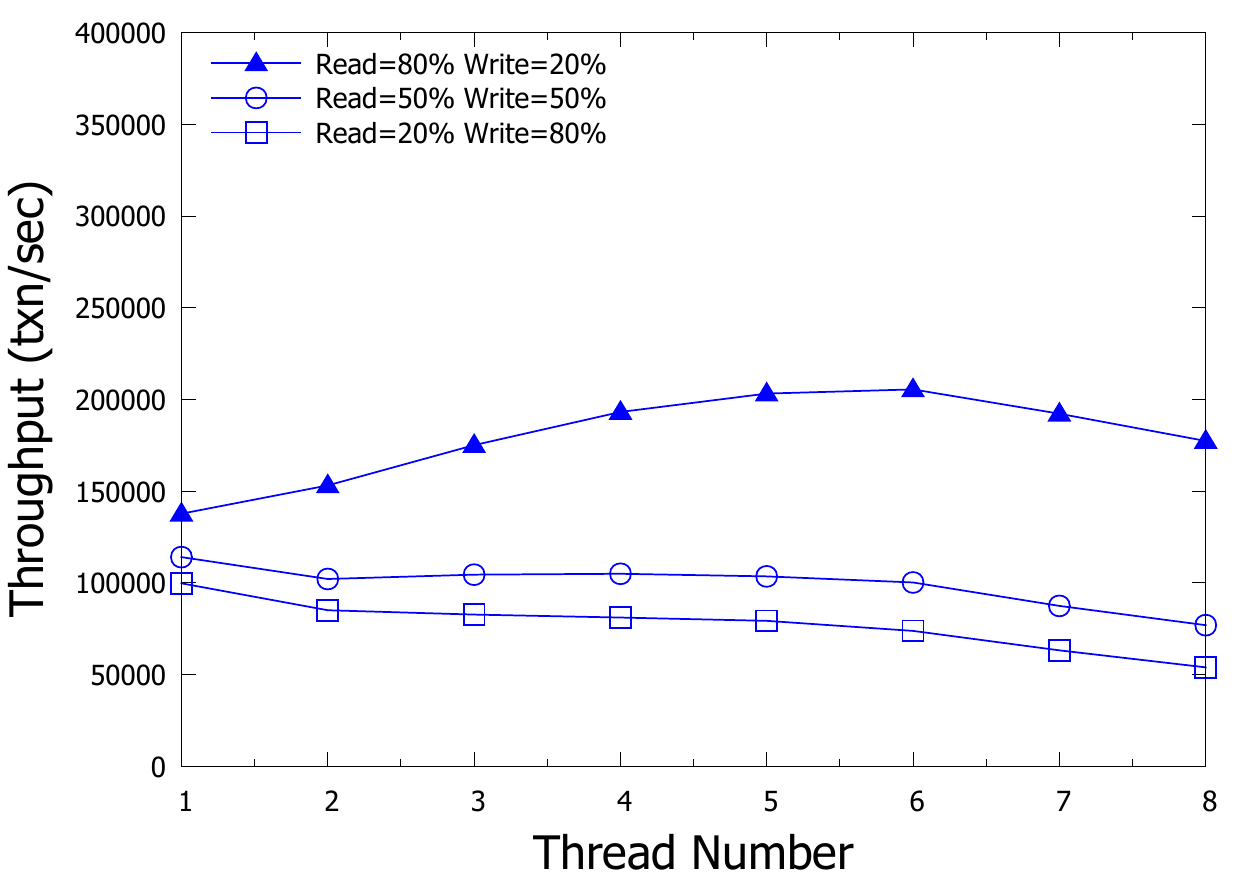}}
  \subfigure[OCC]{
    \label{fig:wr:occ} 
    \includegraphics[width=0.235\linewidth,height=0.22\linewidth]{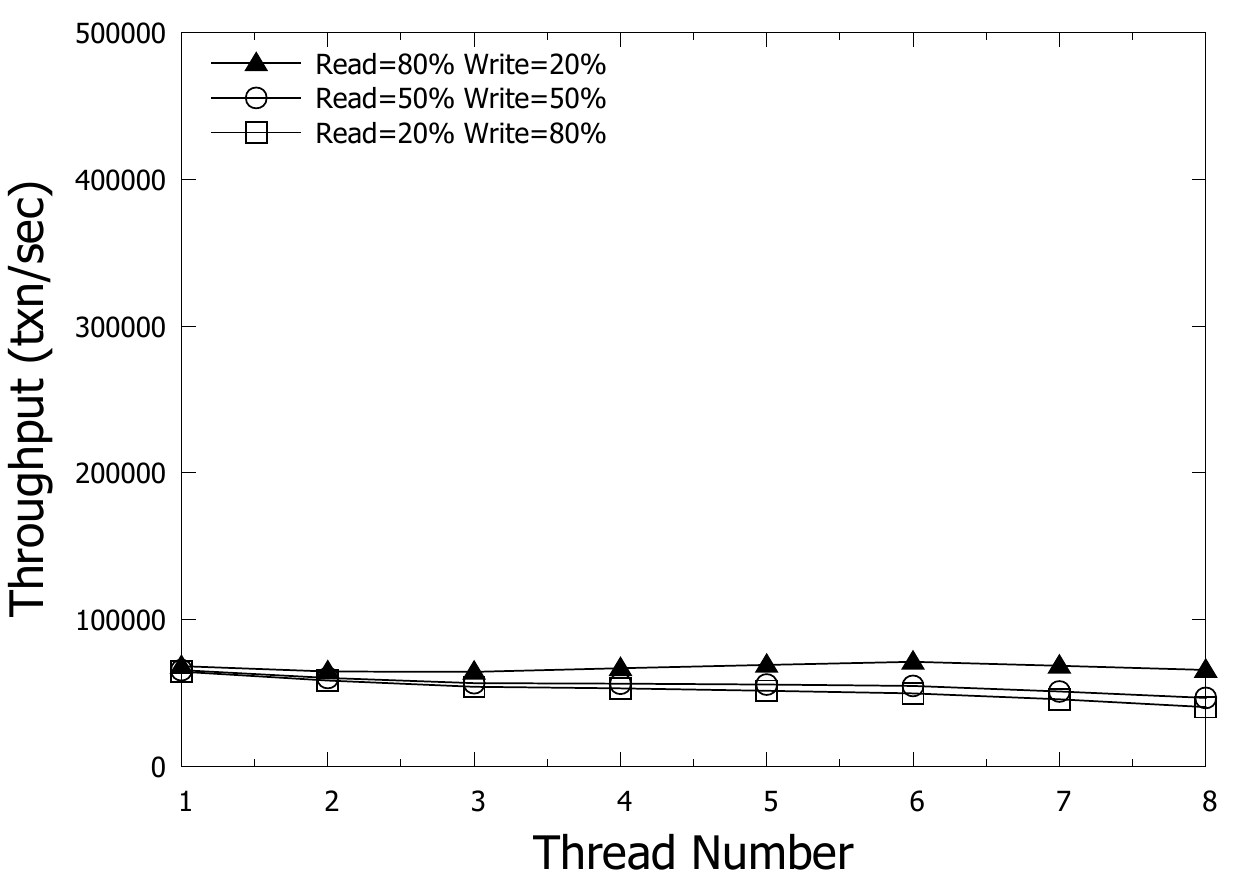}}
  \subfigure[MVCC]{
     \label{fig:wr:mvcc} 
     \includegraphics[width=0.235\linewidth,height=0.22\linewidth]{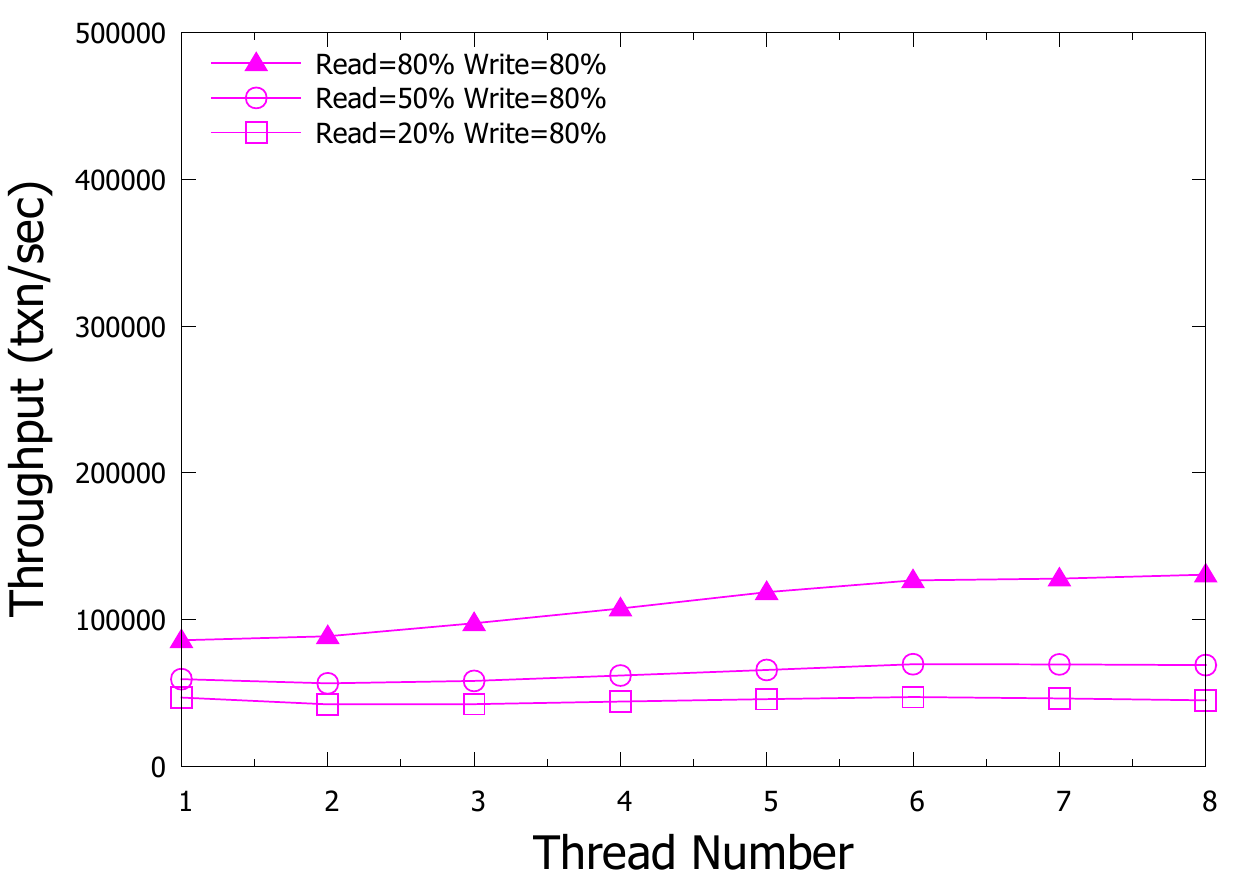}}
  \subfigure[DGCC]{
     \label{fig:wr:dgcc} 
     \includegraphics[width=0.235\linewidth,height=0.22\linewidth]{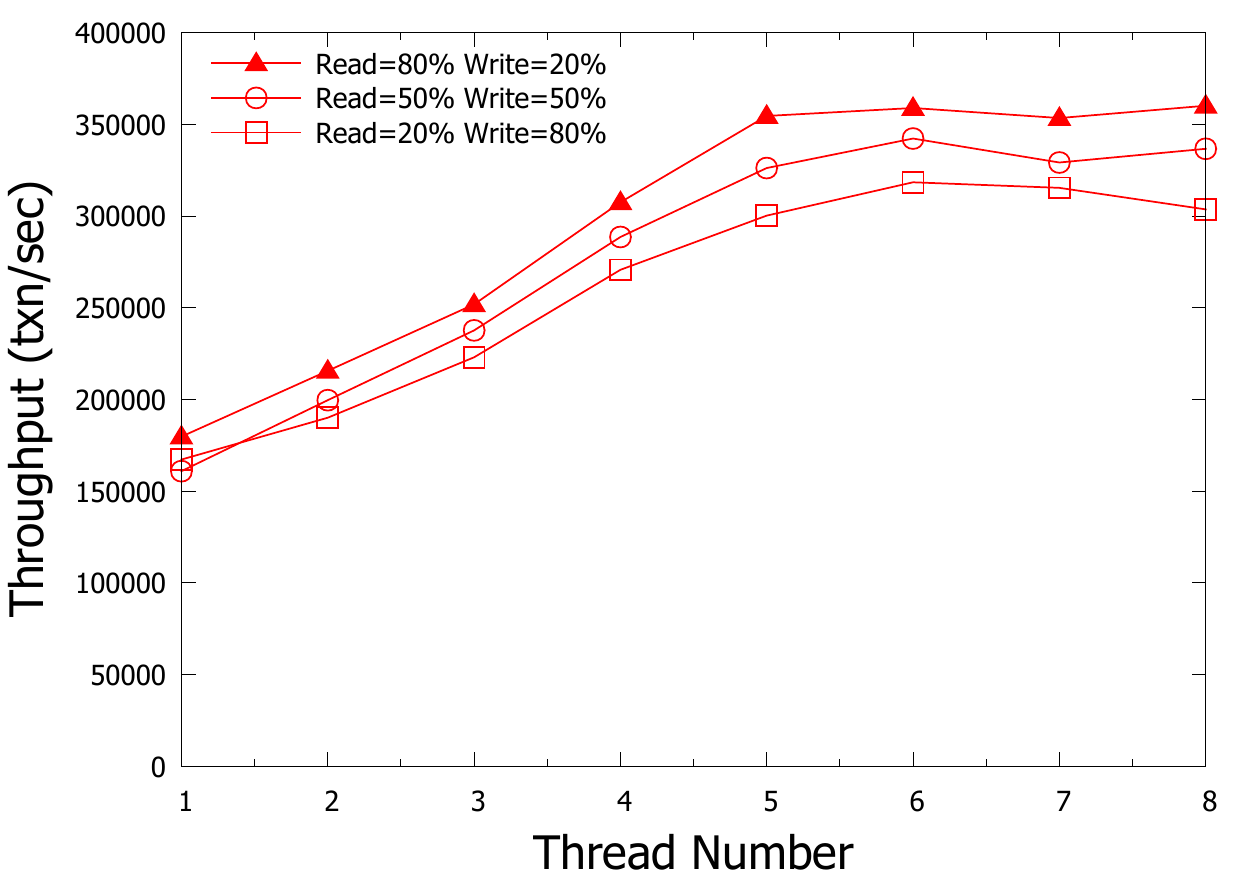}}
  \caption{Effect of Write Operations,$\theta$=0.8}
  \label{fig:wr}
\end{figure*}

\subsubsection{Transaction Logs}
Before a transaction commits,
the system will first generate its log records and
flush them to the log files on disk.
The recovery component has logger threads which are responsible for flushing the logs to disks.
Traditionally, there are two kinds of logging strategies,
ARIES logging~\cite{mohan1992aries} and Command logging~\cite{malviya2014rethinking}.

In our system,
transactions of one graph commit at the same time.
Instead of generating log records for a single transaction,
our system constructs log records for all the transactions
in a batch simultaneously.
Writing all these log records at the same time
fully utilizes the disk I/O bandwidth,
thereby improving the system's overall performance.

Each vertex in the dependency graph has one log record consisting of
the vertex's function ID, parameters, and dependency information.
This information is sufficient for the reconstruction of the
dependency graph during recovery.
Our logging scheme combines the advantages of both ARIES and command logging.
No real data needs to be recorded in the log files,
hence reducing the size of the logs.
During recovery,
we only need to replay the log records to reconstruct the dependency graphs and then execute the reconstructed graph.

\subsubsection{Checkpointing}
In order to recover our database within a bounded time,
our system takes periodic checkpointing.
Our recovery component maintains several checkpointing threads.
The entire memory is divided up into sections and
each checkpointing thread is responsible for one such section.

Even as the checkpointing threads are working,
transactions continue to execute.
However, those commits are not reflected in the checkpointing.
This means our checkpointing is not a consistent snapshot of the database, and
it needs to combine with the transaction logs.

To recover from a failure,
our system first reloads the latest checkpoint and replays the transaction log records from that time point.
It then reprocesses the committed transactions.

\subsection{Storage Manager}

The system's storage manager is designed to maintain the whole data in the database.
It interacts with the execution engine to retrieve/insert/update/delete the data.
Both the B$^+$-tree index and hash index are supported.

DGCC guarantees the serializability and zero-conflict for write and read operations.
However, insert and delete operations also
requires the index to be correctly maintained.
Algorithms~\cite{sewall2011palm,levandoski2013bw} that have been
proposed to exploit more concurrency in indexing are orthogonal to
our proposed concurrency control protocol.
We can make use of any one of them together
with DGCC to enhance the system's overall performance.

Our system maintains all of its allocated memory space on its own to
avoid frequent invocations of system calls (such as malloc).
To eliminate the bottlenecks in the storage manager,
the system divides up the pre-allocated memory space,
and assigns a worker thread to each section to insert/delete its data.
It usually has many insert/delete operations for OLTP applications.
The memory usage efficiency should be taken into consideration.
A garbage collection thread in the storage manager will be invoked periodically
to collect inactive objects and compact the memory space.

\subsection{Statistics Manager}

As shown in Figure~\ref{fig:architecture},
our system has a statistics manager
that collects runtime statistics information (such as real-time throughput, latency etc.).
It also interacts with the other components
to adjust the system configuration dynamically.
For example, since our system processes transactions in batches due to DGCC,
the size of the dependency graph affects both the throughput and latency.
A larger batch size is better for supporting higher 
throughput, and a smaller batch size provides
a faster response time.
The maximal batch size can be adjusted accordingly based on the
statistics and the requirements.
Furthermore, using the statistics information collecting from the storage manager,
the system decides when to invoke the garbage collection thread.

\section{Experiments}
In this section, we evaluate the effectiveness of DGCC,
by comparing it with the following concurrency control protocols,
which implemented in a multicore DMBMs~\cite{yustaring}.
\begin{itemize}
\item \textbf{2PL} - Two-Phase Locking with deadlock detection
\item \textbf{OCC} - Optimistic Concurrency Control,
\item \textbf{MVCC} - Multi-Version Concurrency Control
\item \textbf{DGCC} - Dependency Graph based Concurrency Control
\end{itemize}

In our evaluation, general optimizations for 2PL, OCC and MVCC are
enabled to make a fair comparison. They are optimized with a
customized memory allocation component to avoid malloc syscall.
Moreover, instead of centralized lock tables, all of them support
decentralized record-level lock tables.

All the experimental evaluations are conducted on a server with
Intel Xeon 2.2~GHz 24-core CPU with 48 hyperthreading and
64GB RAM. It contains 4 NUMA nodes. To eliminate
the effects of NUMA architecture, we run most experiments
in one NUMA node with 6 cores. Each core has a private 32KB
L1 cache, 256KB L2 cache and supports two hyper threads.
The cores in the same NUMA node share a 12MB L3 cache.

We use two popular OLTP benchmarks, namely YCSB~\cite{cooper2010benchmarking} and TPC-C~\cite{tpcc1111}.
YCSB is used to evaluate the performance of these concurrency control protocols
under different contention rates caused by data access skewness.
TPC-C is used to simulate a complete order-entry environment
whose transaction scenario is much more complex than that of YCSB.
The contention rate in TPC-C is controlled by the number of warehouses.

The main purpose of concurrency control protocols is to resolve contentions in
a multi-programmed environment.
There are three factors that typically dominate the intensity of the contentions.
The first one is the ratio of write operations in the workload.
The second is the data access skewness, in particularly, frequently accessed data
encounter contention more easily.
Another factor is the number of concurrent worker threads.
The higher the number of parallel transactions, the larger is the
probability of contention.

In the following experiments, we evaluate the performance of DGCC with respect to
all these factors using the two benchmarks.
\yaochang{
The parameters we used in the experiments are listed in Table \ref{table:exp_para}, with default setting underlined.}
\begin{table}[tb]\scriptsize
\centering
\caption{Parameter Ranges for Evaluations}
\begin{tabular}{|c|c|c|} \hline
\textbf{Parameter} & \textbf{Description} & \textbf{Range} \\ \hline
$\theta$ & YCSB Zipfian parameter& \tabincell{c}{0.0, 0.5, 0.6,0.7,\underline{0.8}} \\ \hline
$\gamma$ & YCSB read/write ratio& \tabincell{c}{4,\underline{1},0.25}\\ \hline
$\kappa$ &worker thread number& \tabincell{c}{1,2,3,4,5,6,7,\underline{8}} \\ \hline
$\delta$ & \tabincell{c}{maximal batch size}& \tabincell{c}{100,300,500,800,\underline{1000},\\ 5000,10000,20000} \\ \hline
\end{tabular}
\label{table:exp_para}
\end{table}

\begin{figure*}
\centering
\begin{minipage}[c]{\textwidth}
\centering
    \subfigure[YCSB Low Contention,$\theta$=0.5]{
       \label{fig:ycsb_thr:low} 
       \includegraphics[width=0.31\linewidth,height=0.22\linewidth]{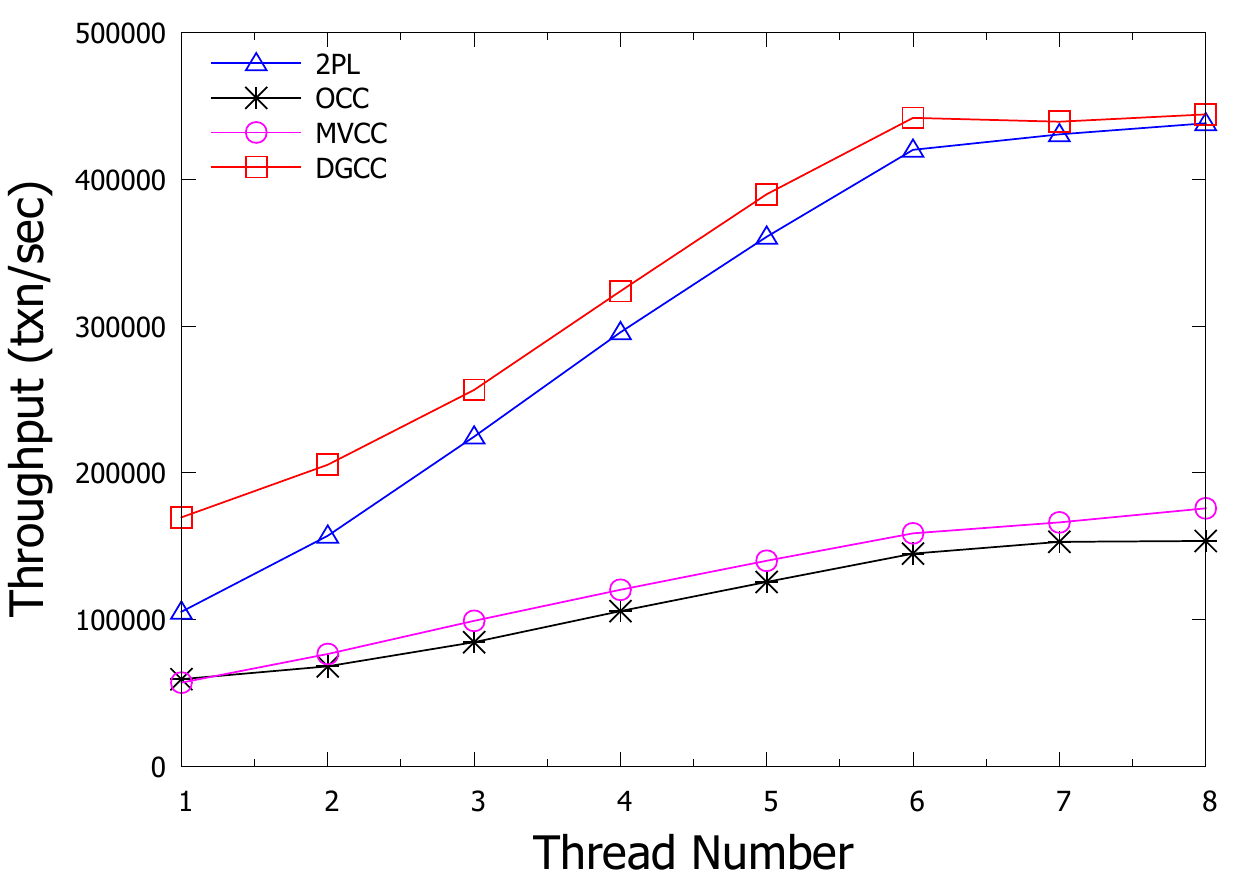}}
     \subfigure[YCSB Median Contention,$\theta$=0.7]{
       \label{fig:ycsb_thr:mid} 
       \includegraphics[width=0.31\linewidth,height=0.22\linewidth]{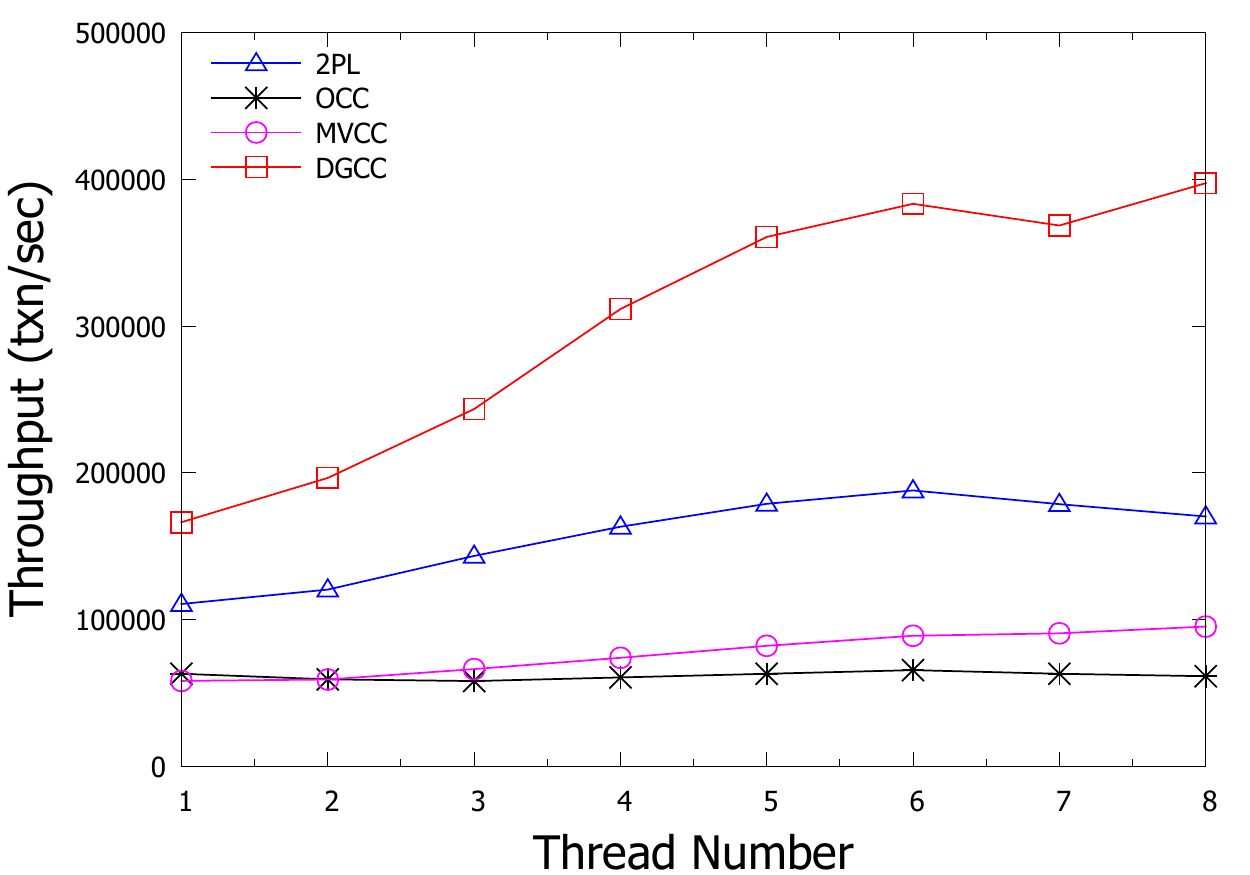}}
     \subfigure[YCSB High Contention,$\theta$=0.8]{
        \label{fig:ycsb_thr:high} 
        \includegraphics[width=0.31\linewidth,height=0.22\linewidth]{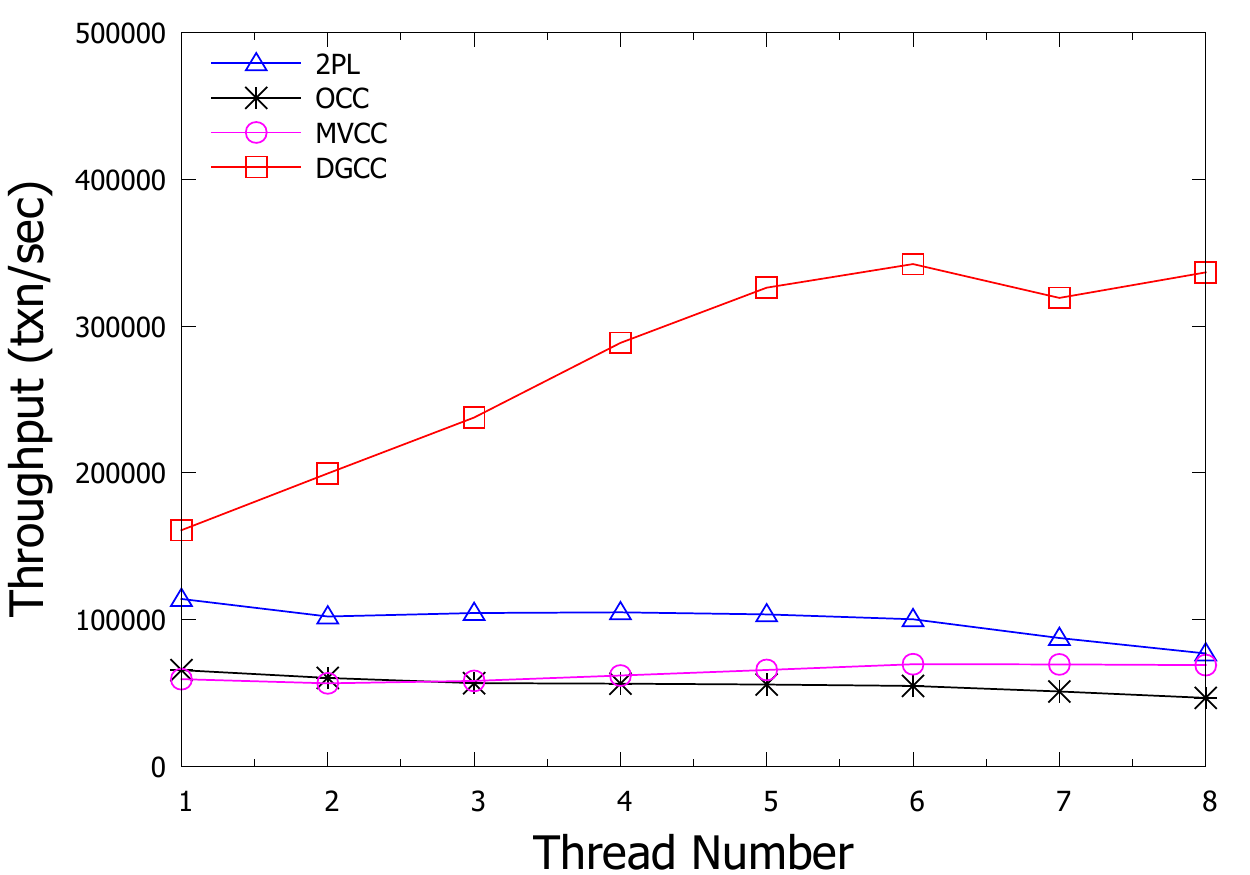}}
     \vspace{-.15in}
     \caption{Throughput for the YCSB workload,$\gamma$=1}
     \label{fig:ycsb_thr}
\end{minipage}
\begin{minipage}[c]{\textwidth}
\centering
    \subfigure[TPC-C]{
       \label{fig:tpcc_thr:tpcc} 
       \includegraphics[width=0.31\linewidth,height=0.22\linewidth]{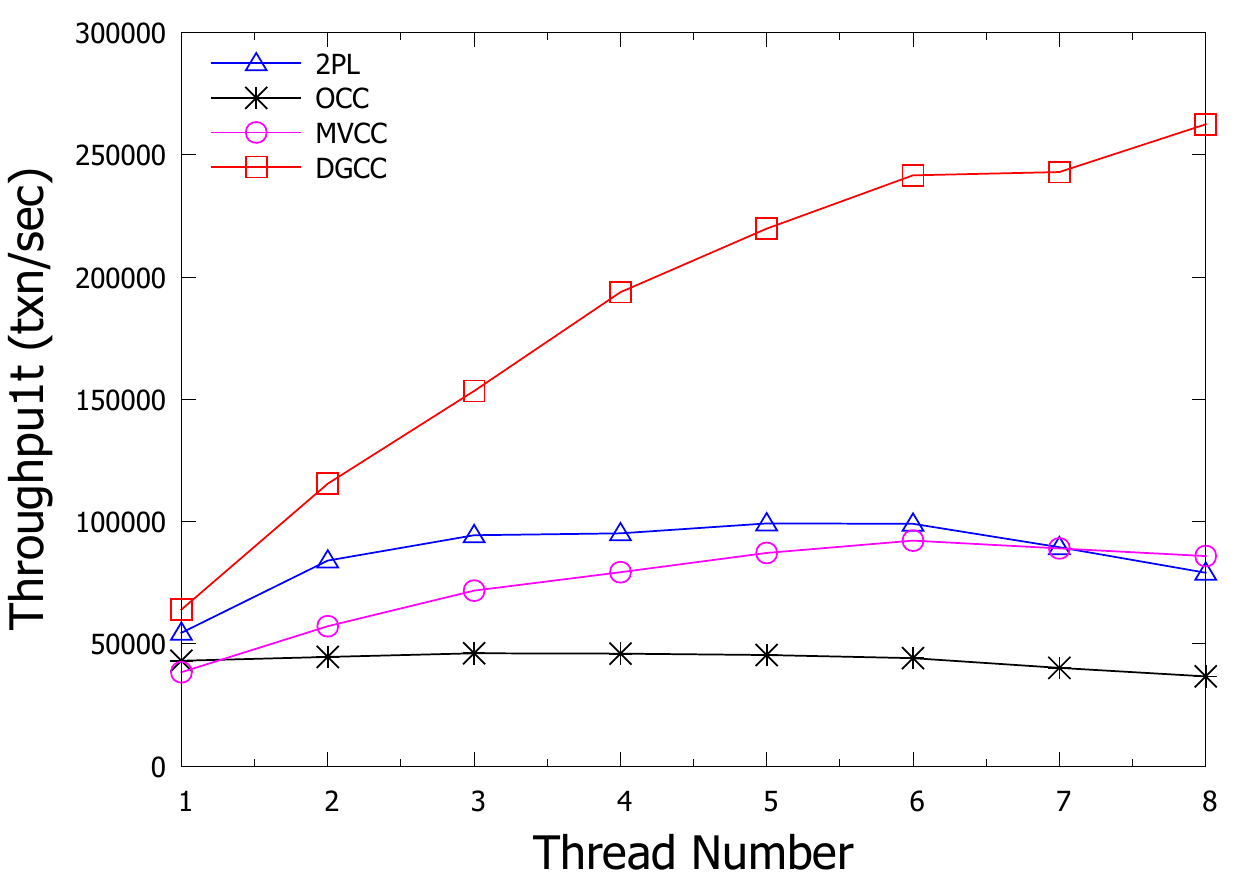}}
     \subfigure[NewOrder]{
       \label{fig:tpcc_thr:neworder} 
       \includegraphics[width=0.31\linewidth,height=0.22\linewidth]{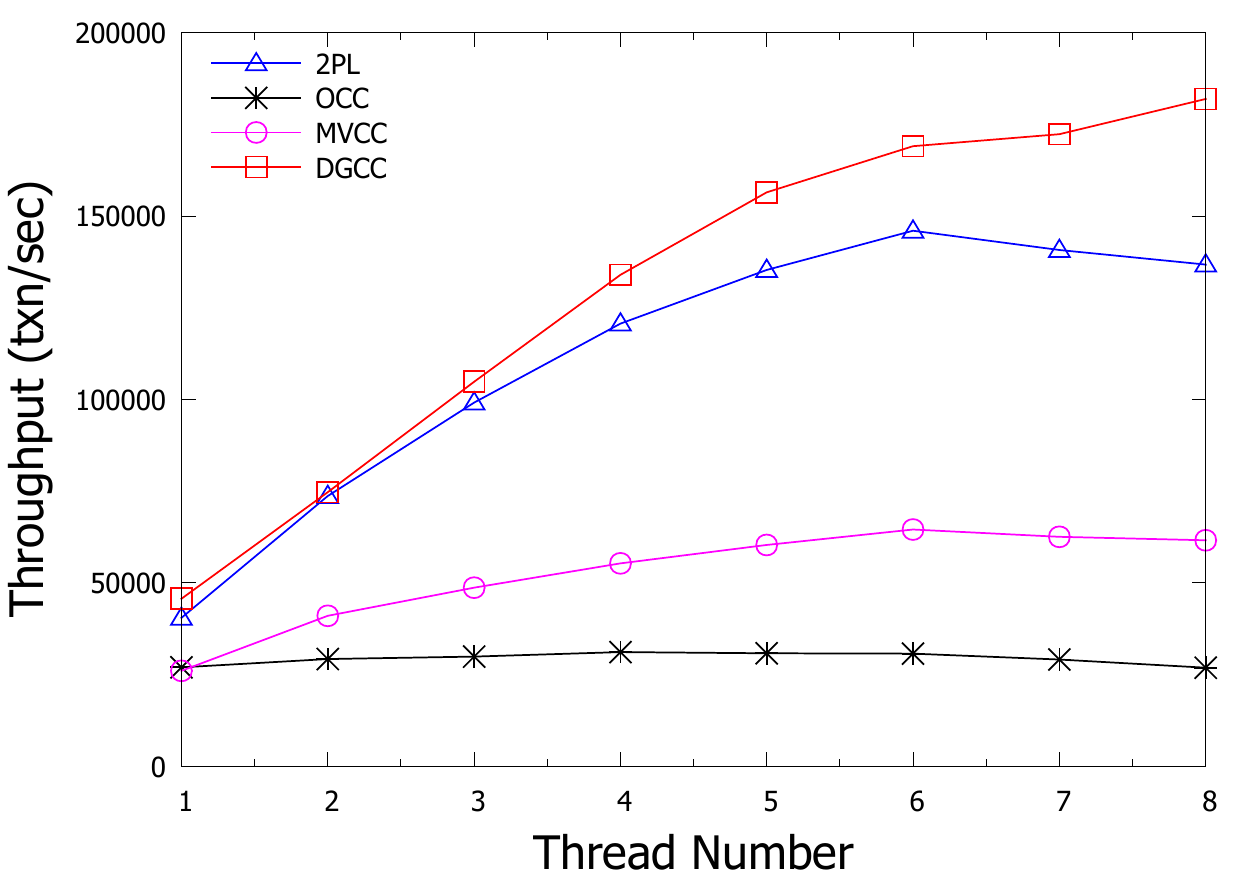}}
     \subfigure[Payment]{
        \label{fig:tpcc_thr:payment} 
        \includegraphics[width=0.31\linewidth,height=0.22\linewidth]{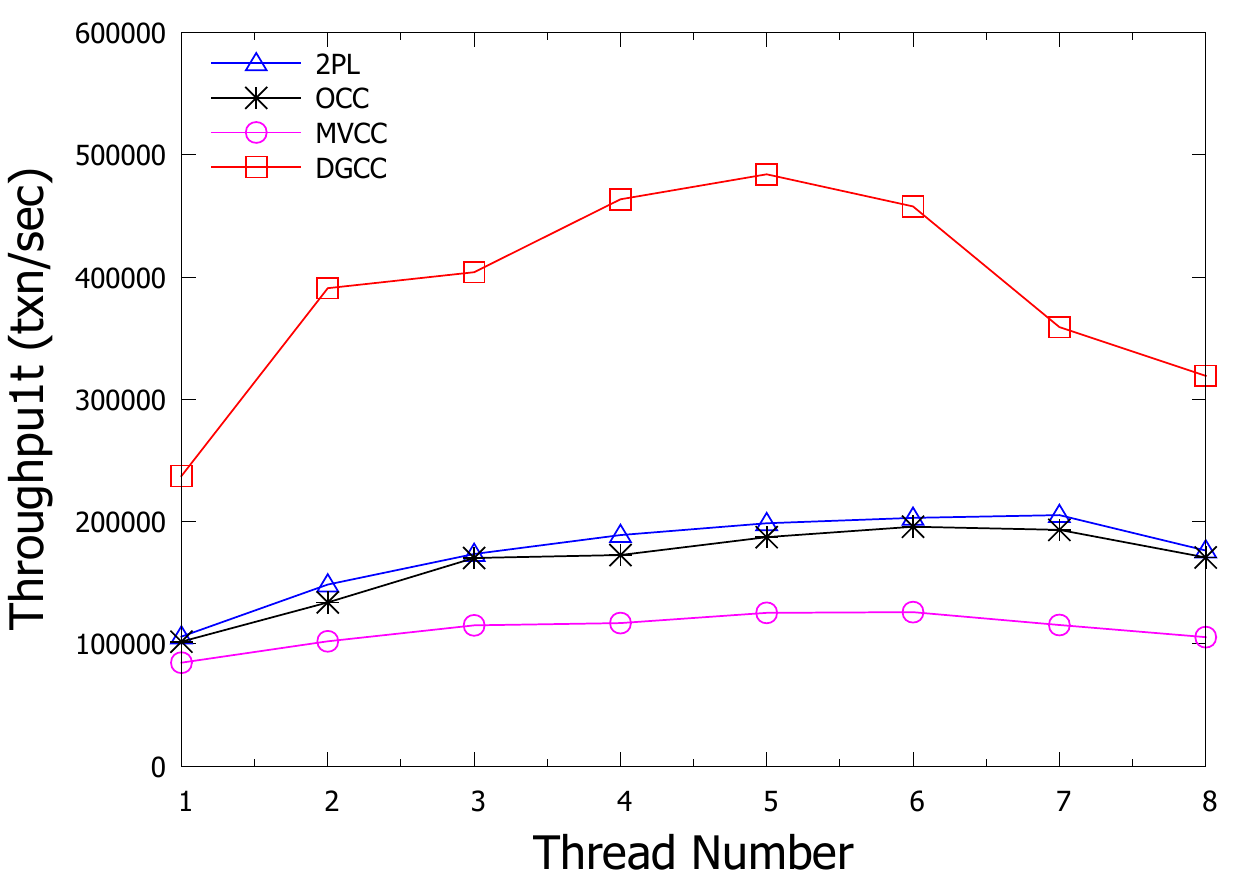}}
     \vspace{-.15in}
     \caption{Throughput for the TPC-C workload}
     \label{fig:tpcc_thr}
\end{minipage}
\end{figure*}

\subsection{Read vs Write Intensive Workloads}

Since read-only transaction pieces will not generate any contentions,
we used the YCSB benchmark that has both read and write pieces.

Figure \ref{fig:wr} shows the performance of different concurrency control protocols on
three workloads of different read/write ratios.
All protocols perform better on workload of more read pieces.
As the write ratio increases in the workload, the performance
of 2PL, OCC and MVCC drops dramatically.
Since more write pieces translate to higher probabilities of contention,
2PL, OCC and MVCC need to spend a lot of time resolving
contentions.
DGCC is significantly more resilient to this increase.
There is little difference between reads or writes at the
dependency graph construction phase.
The performance reduction in DGCC is due to the fact that
write pieces usually take more time than read pieces.

\subsection{Scalability}
In Figure \ref{fig:ycsb_thr}, we test the performance of the
four concurrency control protocols
under different contention rates.
Contention rate is controlled by setting the parameter $\theta$
in YCSB's Zipfian distribution.
The read/write ratio $\gamma$ in the experiments are fixed to 1.

In summary,
DGCC shows the best performance under different contention rates.
The benefits come mainly from the separation of contention resolution and execution.
By resolving contentions in advance, no worker thread is blocked during the execution
and the acyclicity of the dependency graph avoids the aborts caused by contention.

It is notable that in Figure~\ref{fig:ycsb_thr:low}, 2PL has a comparable performance with DGCC.
In this experiment, $\theta$=0.5 in Zipfian distribution results
in the lowest contention rate.
Under this scenario, 2PL has little overhead because it does not
waste time on acquiring locks.
Further, deadlocks rarely occur because the probability that
more than one transaction competing for the same data is low.

The reason for the drop in DGCC's performance when
the thread count is increased
(7 and 8 in our experiment) are two fold.
Firstly, our experiments ran on 6 cores.
When more than 6 threads are running concurrently,
the overhead of context switch becomes significant.
Secondly, the increase in thread count will inevitably result in more contention,
resulting in higher overhead to resolve contention for all four protocols.

OCC and MVCC are timestamp based protocols.
When the data access distribution is not very skewed,
they scale well with the number of worker threads.
However, compared with 2PL and DGCC, timestamp based protocols
have to spend time in assigning the timestamp to each transaction.
In order to guarantee the correct serial order,
such systems usually use a centralized component to perform the assignment
and this easily contributes to the performance bottleneck.
Moreover, at the commit time, OCC and MVCC
must validate that the execution is serialized according to the
assigned timestamps.
Whether or not a transaction aborts,
transactions accessing the same data have to be validated one by one.

The main cost of OCC and MVCC comes from processing abort
when the contention is high.
Unlike aborts in lock based protocols,
aborts at the commit phase not only cost the usual processing time but also
require extra effort in eliminating the
effects of those aborted transactions in the database.

Figure~\ref{fig:tpcc_thr} shows the evaluation of the four protocols using the TPC-C benchmark.
The contention rate in TPC-C is usually controlled by the number of warehouses.
In this experiment,
we set the number of warehouse to 1 so as to create enough contention.
There are five types of transactions in TPC-C: New-Order, Payment, Delivery, Order-Status
and Stock-Level.
New-Order and Payment are the most frequent transactions,
accounting for almost 90\% of the whole benchmark.
Therefore, in addition to the entire benchmark,
we also compare performance using only these two transactions
separately.

\begin{figure*}
\centering
\begin{minipage}[c]{\textwidth}
\centering
      \subfigure[YCSB Low Contention,$\theta$=0.5]{
        \label{fig:ycsb_lat:low} 
        \includegraphics[width=0.31\linewidth,height=0.22\linewidth]{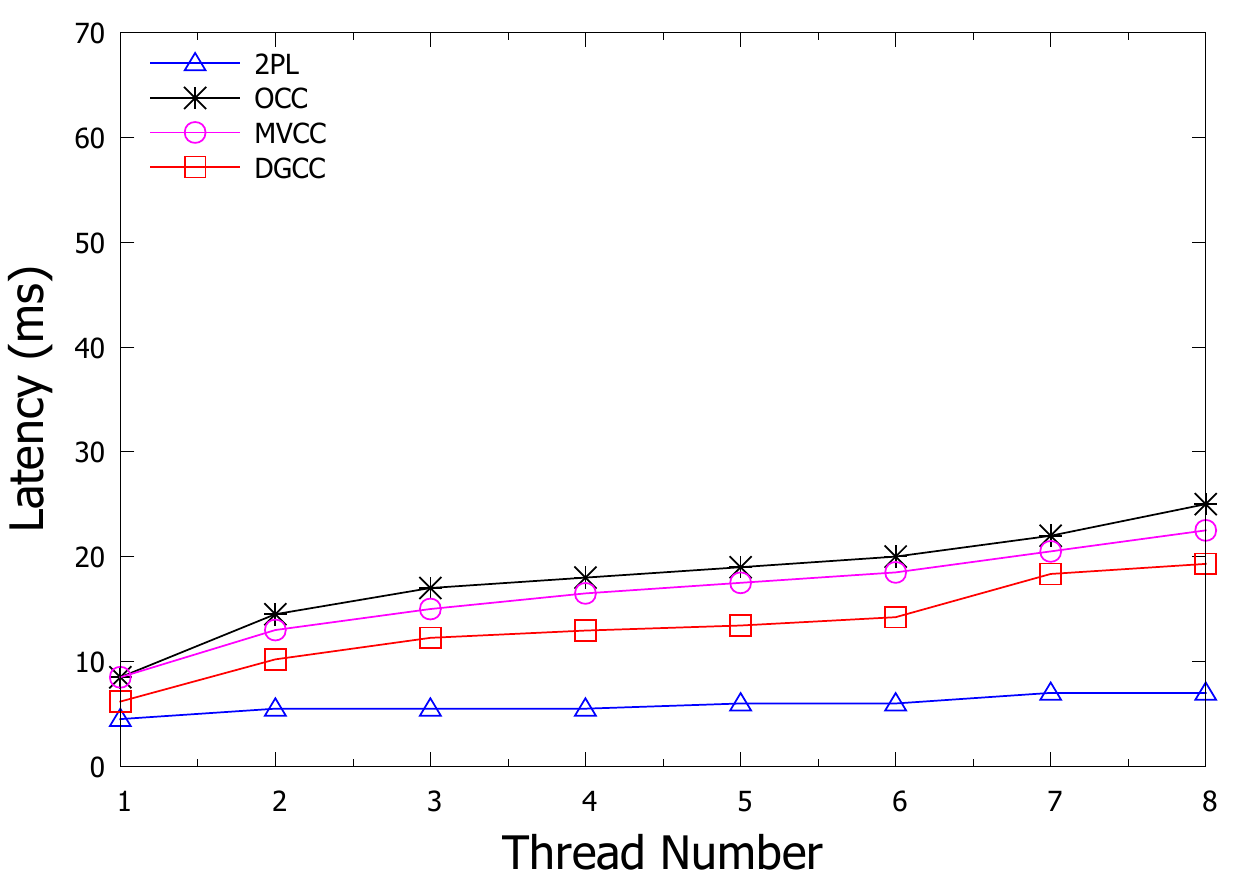}}
      \subfigure[YCSB Median Contention,$\theta$=0.7]{
        \label{fig:ycsb_lat:mid} 
        \includegraphics[width=0.31\linewidth,height=0.22\linewidth]{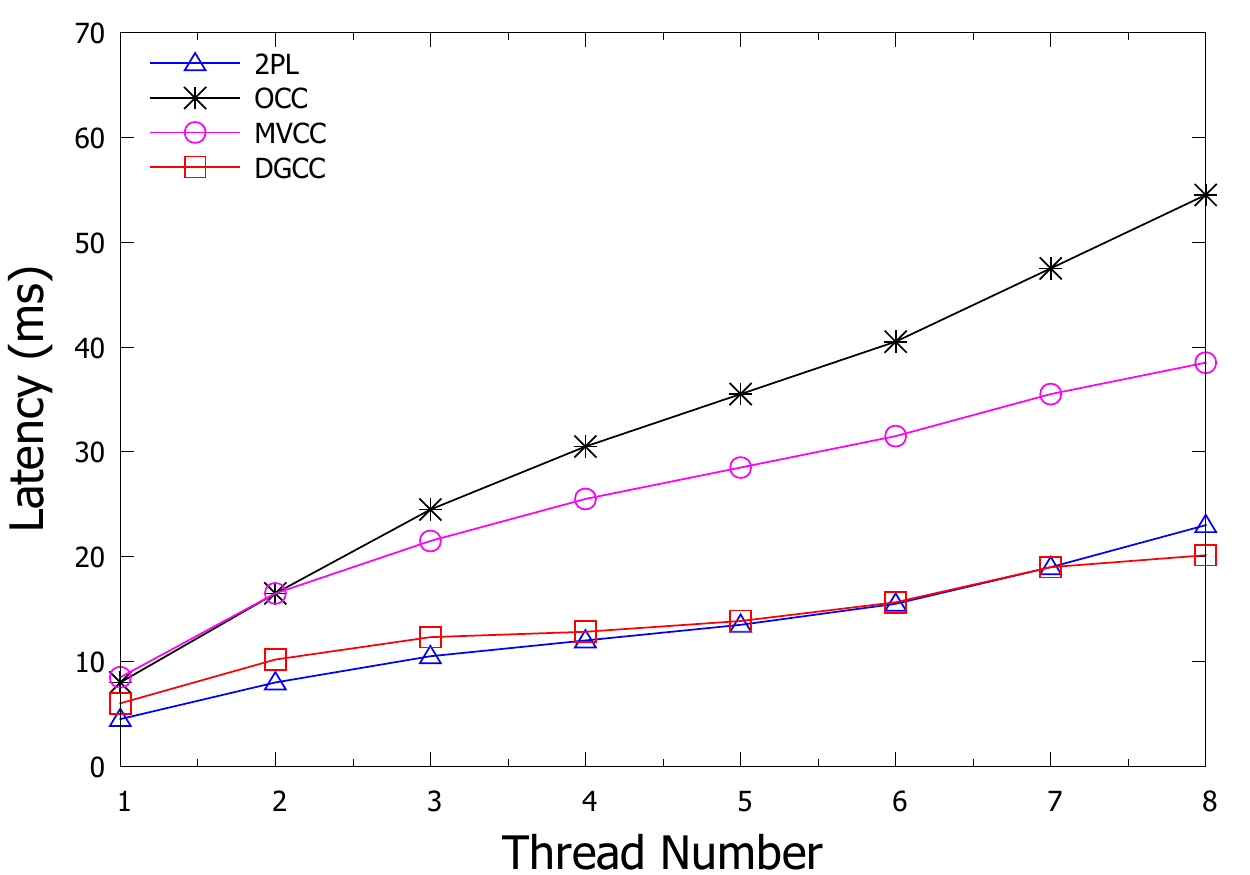}}
      \subfigure[YCSB High Contention,$\theta$=0.8]{
         \label{fig:ycsb_lat:high} 
         \includegraphics[width=0.31\linewidth,height=0.22\linewidth]{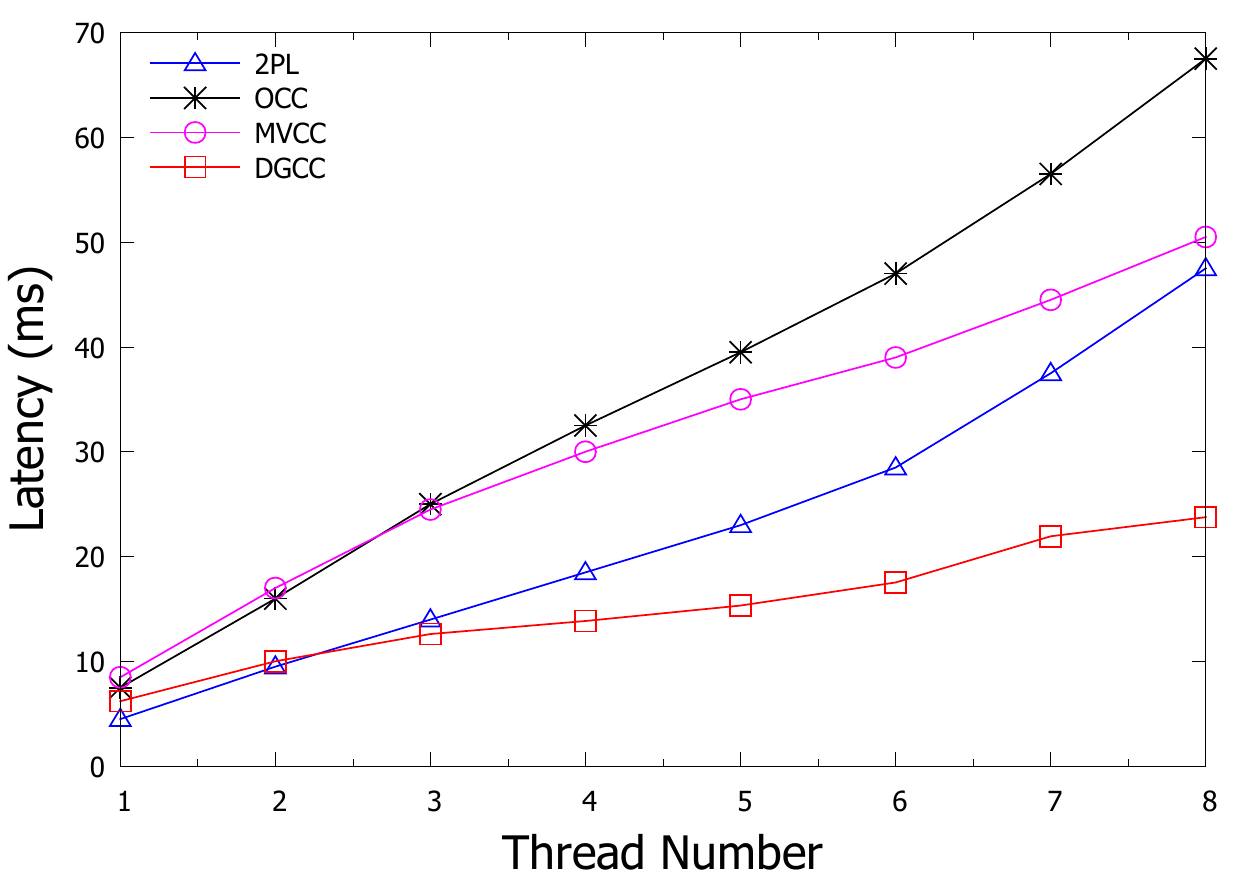}}
      \vspace{-.15in}
      \caption{Average Latency for the YCSB workload, $\gamma$=1}
      \label{fig:ycsb_lat}
\end{minipage}

\begin{minipage}[c]{\textwidth}
\centering
    \subfigure[TPC-C]{
       \label{fig:tpcc_lat:tpcc} 
       \includegraphics[width=0.31\linewidth,height=0.22\linewidth]{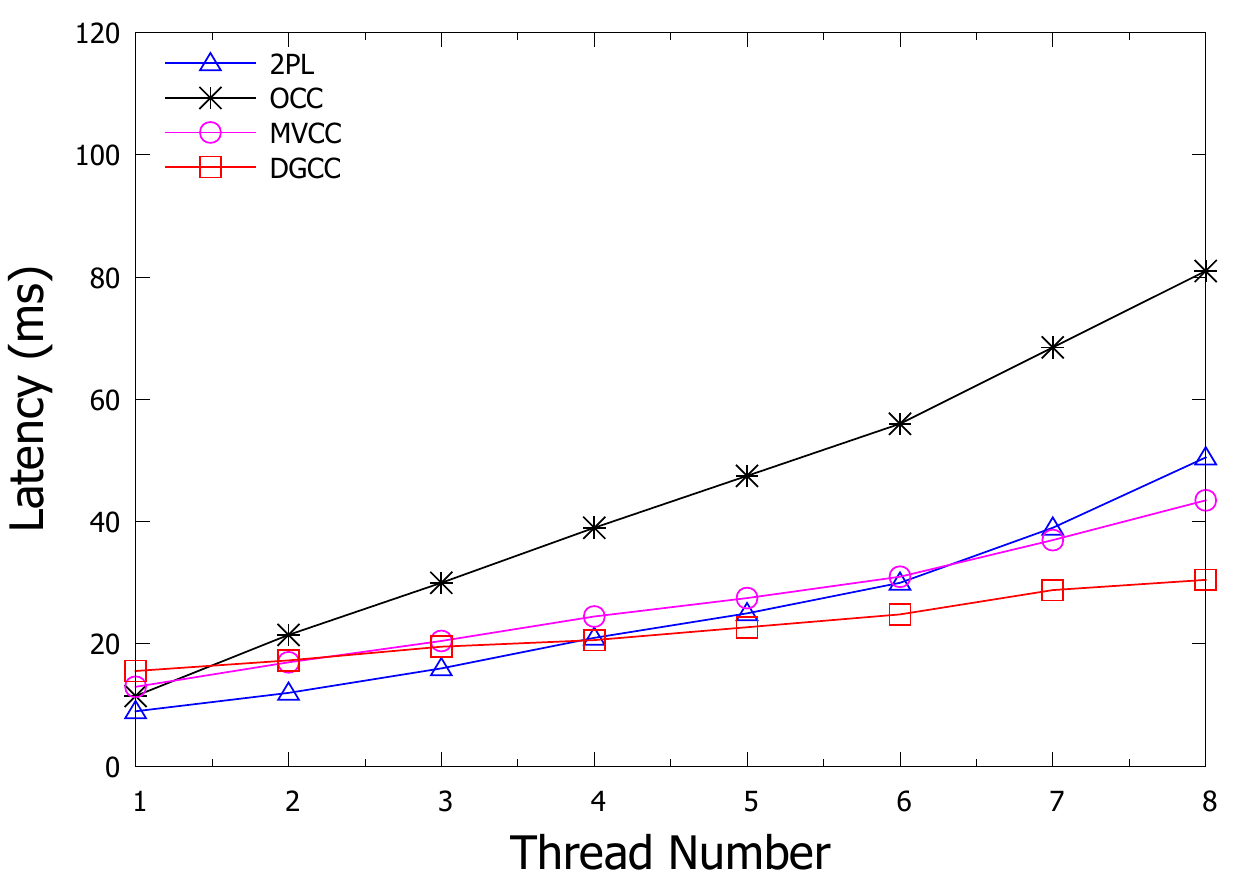}}
     \subfigure[NewOrder]{
       \label{fig:tpcc_lat:neworder} 
       \includegraphics[width=0.31\linewidth,height=0.22\linewidth]{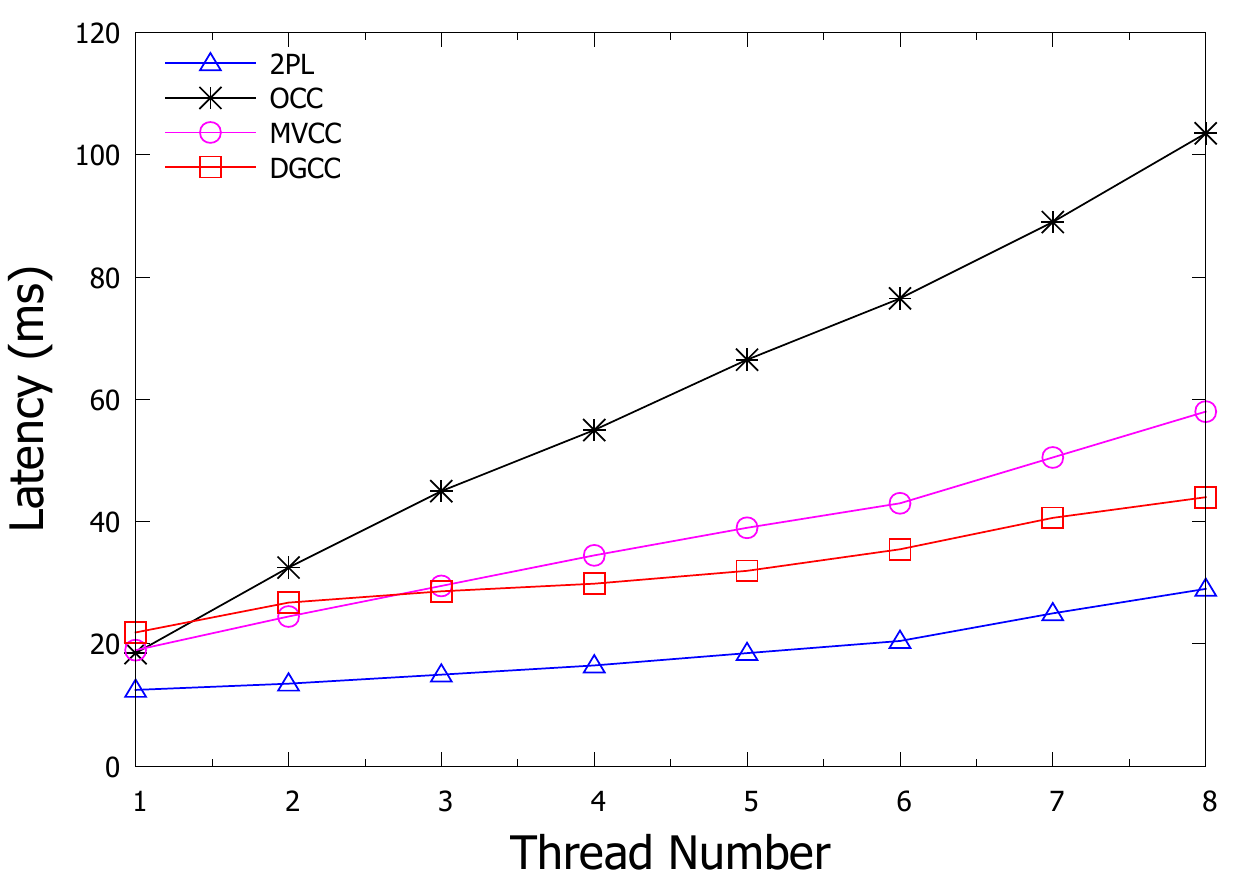}}
     \subfigure[Payment]{
        \label{fig:tpcc_lat:payment} 
        \includegraphics[width=0.31\linewidth,height=0.22\linewidth]{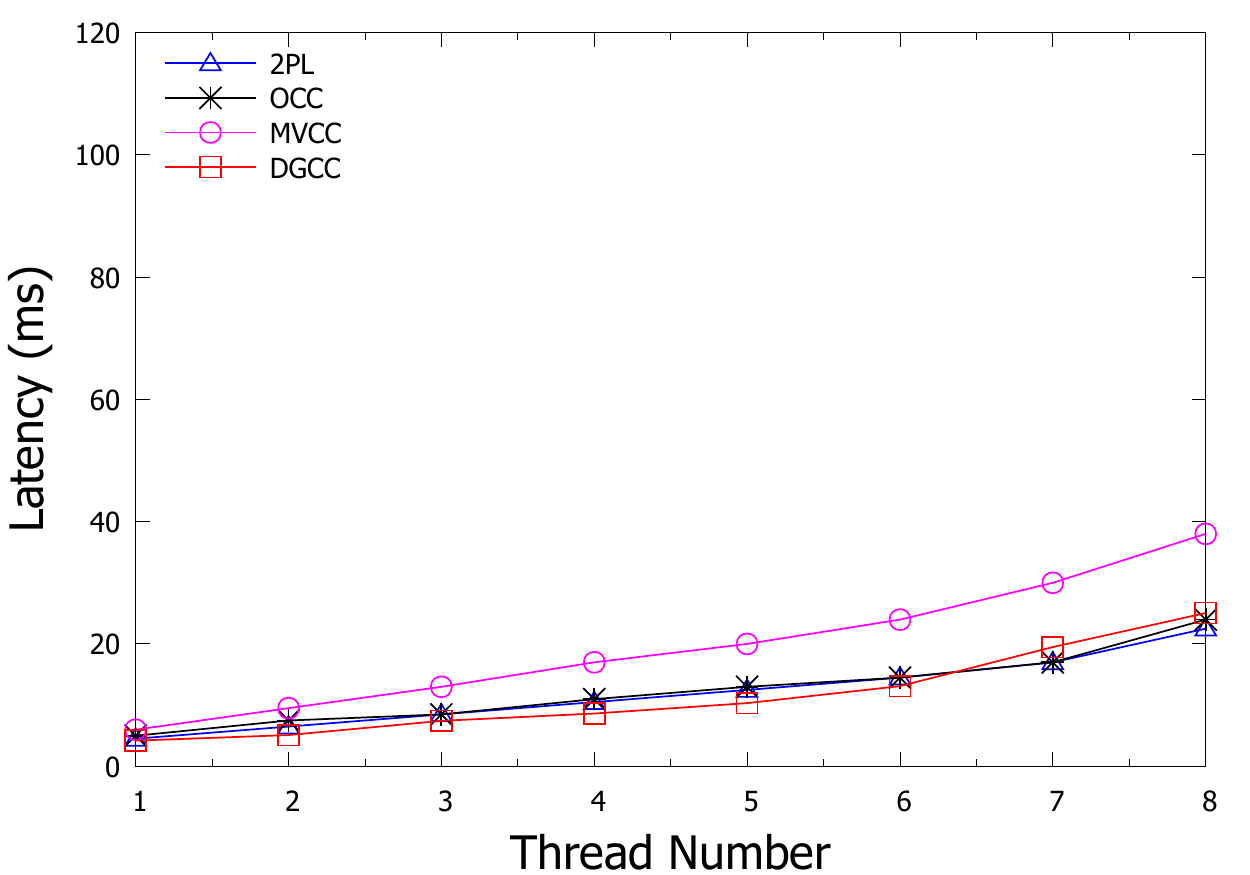}}
     \vspace{-.15in}
     \caption{Average Latency for the TPC-C workload}
     \label{fig:tpcc_lat}
\end{minipage}

\end{figure*}

Figure \ref{fig:tpcc_lat:neworder} shows the results when only New-Order is
considered.
Each New-Order transaction, on average, comprises of ten different items.
These items' information have to be read and the related stock information need to
be updated.
Which item gets accessed is entirely random, and this
leads to a relatively low level of contention.
Results shown in Figure~\ref{fig:tpcc_thr:neworder} are within the expectation.
Although DGCC still achieves the best performance,
2PL comes in a close second.

Figure~\ref{fig:tpcc_thr:payment} shows the situation when only Payment transactions
are involved.
Each Payment transaction tries to record a payment from a customer,
and it needs to update the warehouse.
Those transaction pieces have to be done serially, thereby severely restricting the
inherent parallelism.
Further more, the longer serial execution logic needs more iterations in
the DGCC's execution phase.
This translates to a higher overhead in areas such as
work dispatch and worker thread scheduling, and affects the scalability of DGCC as a result.

Figure \ref{fig:tpcc_thr:tpcc} shows the results on the complete TPC-C workload.
Other transactions amortized the effects of payment transaction,
DGCC has a more balanced workload at each iteration, making it more scalable.
However, the high contention caused by Payment transactions is still the bottleneck
for the other protocols.

\subsection{Data Access Distribution}
In reality, OLTP applications tend to access certain data more frequently.
For example, in an online shopping scenario,
popular items are accessed more frequently than others.
The distribution of data accesses has a significant impact on the
level of contention.
YCSB assumes that data accesses follow a Zipfian distribution
whose parameter $\theta$ controls the skewness.
For a given number of working threads,
a larger $\theta$ translates to a higher contention.

\begin{figure}[h]
\centering
\includegraphics[width=0.45\textwidth,height=0.33\textwidth]{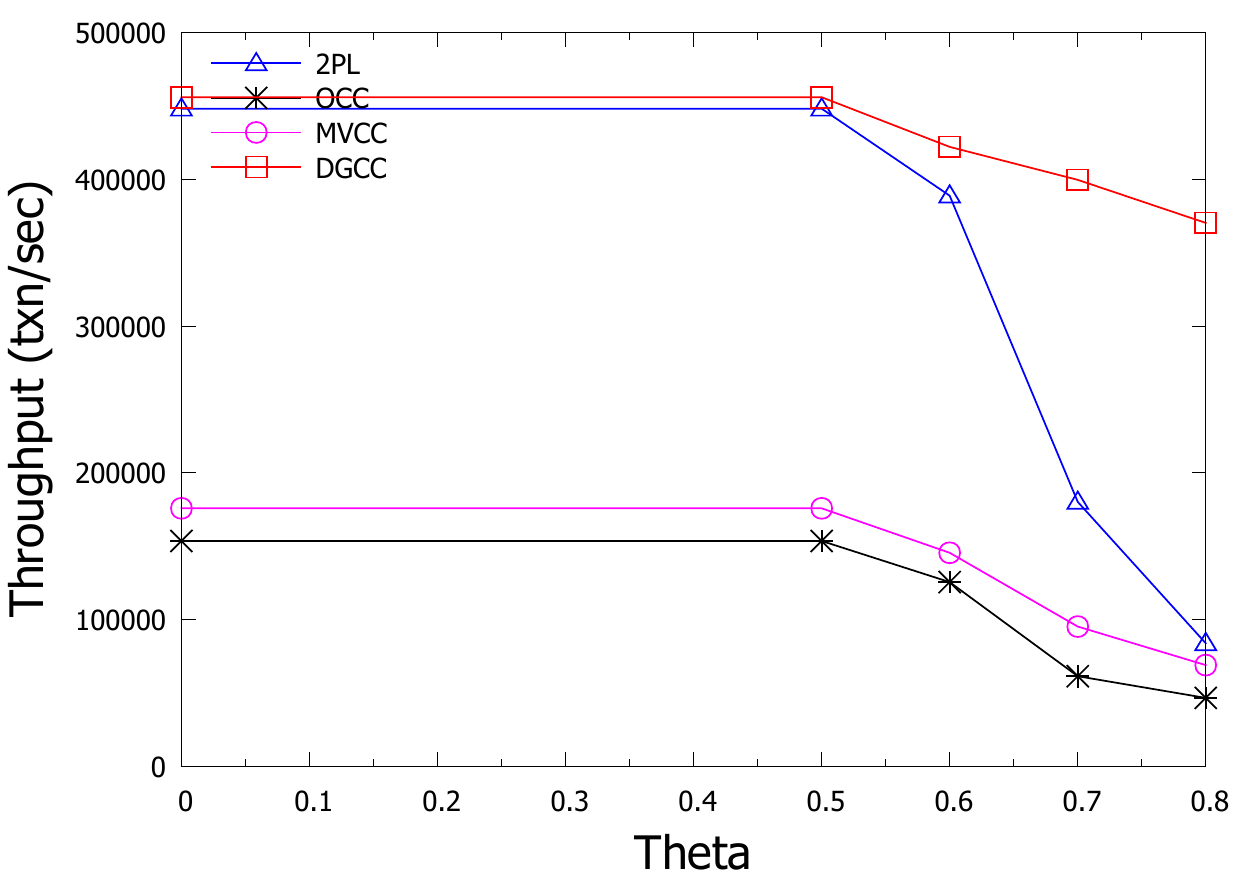}
\caption{Effects of Access Distribution on YCSB, $\kappa=8,\gamma=1$}
\label{fig:ycsb_theta}
\end{figure}

Figure \ref{fig:ycsb_theta} shows the impact of $\theta$ on the
performance of the four protocols.
When $\theta$ is small, data accesses are more likely to be
uniformly distributed. This is the ideal case for all the
protocols.

As $\theta$ increases,
the data access distribution becomes more skewed, resulting
in higher contention and lower performance.
Yet,
compared to 2PL, OCC as well as MVCC, DGCC is
significantly less sensitive to increased contention.
Higher contention may increase the depth of the dependency graph,
and as a result more iterations are required at the execution phase.
With increasing contention,
the concurrently executable work at each iterations tends to decrease.
However, compared to the other protocols,
the overhead incurred by DGCC is lower,
making DGCC more robust to data access skewness.

\subsection{Latency}
%
In this section,
we shall evaluate the latency using the OLTP workloads.
The system maintains a transaction queue to buffer the transactions that have arrived.
The size of the queue affects the average latency of the system.
It also restricts the number of transactions in dependency graphs.
In the experiments,
for each worker thread,
we set the default size of the transaction queue to $1000$.

Figure \ref{fig:ycsb_lat} and \ref{fig:tpcc_lat} show the latency of four protocols
under different workloads.
The average latency of OCC and MVCC increases when
there is more contention.
They require more time to perform validation at the commit phase.
Furthermore, the latency of timestamp based protocols is also affected
by the centralized timestamp assignment.
When there is more contention,
2PL spent much time waiting for locks, leading to an increase in latency.

In both Figure~\ref{fig:ycsb_lat} and \ref{fig:tpcc_lat}, the latency of DGCC
is comparable with the others.
Although DGCC is a protocol that has a
batch processing front end,
the waiting time of one transaction in the transaction queue
is much less than the other protocols.
So the latency of DGCC is actually smaller.
When we take transaction logs into consideration,
DGCC commits a group of transactions at the same time
and the log size is much smaller than traditional ARIES log.
As a result, it invokes less syscalls to flush the
logs to disks, and consequently, can make better use of the I/O bandwidth.
Overall, this resulted in lower latency compared to the others, thus confirming the efficiency of DGCC.

\subsection{Effects of Batch Size}
DGCC first constructs a dependency graph for a batch of transactions.
The batch size is constrained by the number of transactions in transaction queue and
our pre-defined maximal batch size $\delta$.
\yaochang{
In practice, the batch size changes dynamically.
In particular,
when there are more transactions waiting in the transaction queue,
a larger batch size is used.
}

\begin{figure}[bht]
\centering
  \subfigure[Throughput]{
    \label{fig:tpcc_graph_size} 
    \includegraphics[width=0.45\textwidth,height=0.31\textwidth]{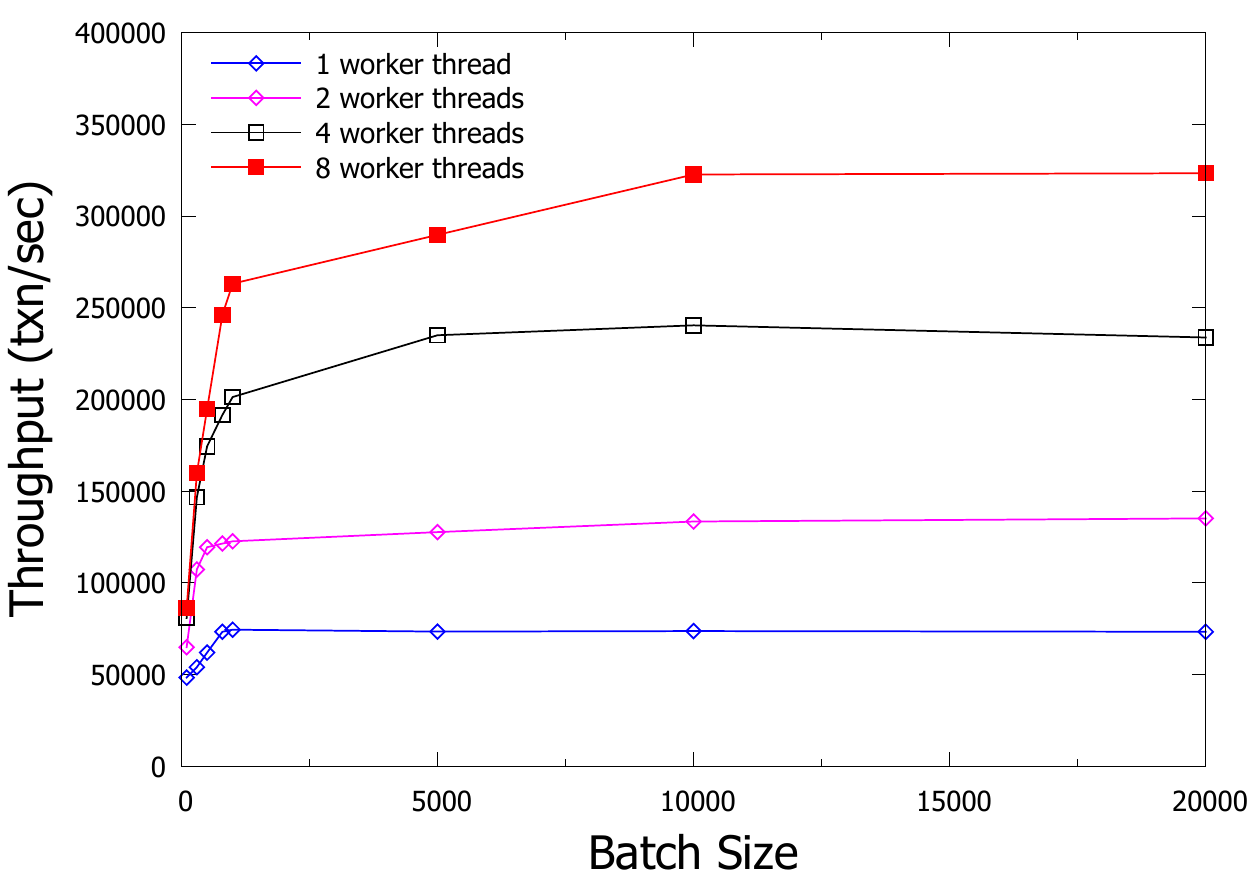}}
  \subfigure[Latency]{
    \label{fig:wr:tpcc_graph_size_lat} 
    \includegraphics[width=0.43\textwidth,height=0.31\textwidth]{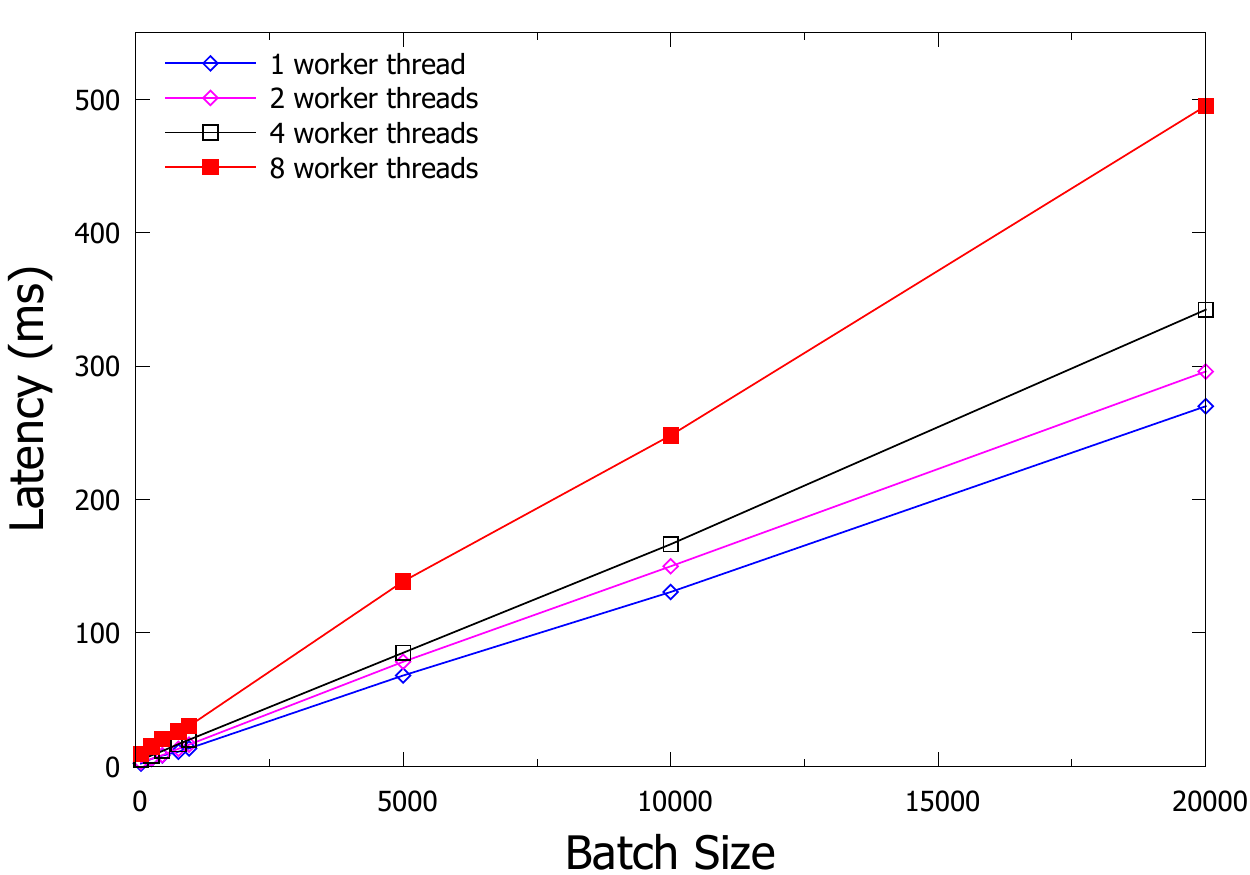}}
  \caption{Effects of Dependency Graph Size on Throughput and Latency}
  \label{fig:tpcc_graph}
\end{figure}

Figure \ref{fig:tpcc_graph_size} shows the effects of the batch size on TPC-C workload.
When the number of worker threads is fixed,
the throughput of the system increases with the batch size.
The increase stops when the computation resource is fully stretched.
From such a point onwards,
a larger dependency graph leads to higher latency.

When there are more worker threads,
it always needs a larger size to fully exploit their computation potential.

\section{Related Work}

Systems with lock based protocols typically require a lock manager, in which
lock tables are maintained to grant and release locks.
The data structure in lock manger is usually very large and complicated,
which incur both storing and processing overheads.

Lightweight Intent Lock(LIL)~\cite{kimura2012efficient} was proposed to maintain
a set of lightweight counters in a global lock table instead of lock queues for intent locks.
It simplifies the data structure of intent locks.
However, the transactions that cannot obtain all the locks
have to be blocked until receiving a release message from other transactions.

%

In order to reduce the cost of a global lock manager,
~\cite{gottemukkala1992locking,lehman1996design} propose to keep lock states with each data record.
However, this idea requires each record to maintain a lock queue, and hence increases
the burden of record management.
By compressing all the lock states at one record into a pair of integers,
~\cite{ren2012lightweight} simplifies the data structure to some extent.
However, it achieves this by dividing the database into disjoint partitions,
which sacrifice its performance and scalability for workload of high contention.

Several in-memory database prototypes that emphasize on scalability
in multi-core systems have been proposed recently.
~\cite{yustaring} implemented an in-memory database prototype and evaluated
the scalability of seven concurrency control methods.
While the reasons differ, the overall result is that
none of the methods can scale beyond 1024 cores.
For lock based methods, lock thrashing and
deadlock avoidance are the main bottlenecks.
For time-stamp based methods, the main issues are the
high abort ratio and the need for a centralized time-stamp allocation.

\cite{kallman2008,pandis2010data,kemper2011hyper} assume
that data in an in-memory database is partitioned,
so as to remove the need for concurrency control.
~\cite{kallman2008} proposes H-STORE, a partitioned database
architecture for in-memory database.
Only one thread in each partition is responsible for processing transactions,
and there is no need for concurrency control within a partition.
DORA~\cite{pandis2010data} is similar to H-STORE, in that
it uses a data partitioning strategy and sends queries to different
partition's worker for processing.
However, unlike H-STORE,
it is able to support concurrency execution of queries in a partition to a certain extent.
Both systems cannot scale well for skewed workload and multi-partition transactions.

Hekaton~\cite{diaconu2013hekaton}, the main memory component of SQL server,
employs lock-free data structures and OCC-based MVCC protocol
to avoid applying writes until the commit time.
However, the centralized timestamp allocation remains the bottleneck, and
the read operations become more expensive, since each read needs to
update other transaction's dependency set.

\cite{tu2013speedy} presented Silo, an in-memory OLTP database prototype
optimized for multi-core systems.
Silo supports a variant of OCC method
which employs batch timestamp allocation to alleviate
the performance loss.
However, workloads with high contention still affect its performance
and scalability.

Transactional memory~\cite{herlihy1993transactional,dice2006transactional}
has been shown to provide scalability with less programming complexity.
Hence, it attracts much attention.
~\cite{leis2014exploiting,wang2014using} exploit hardware transaction memory.
by chopping up transactions into small operations in order to fit them in hardware transaction memory. They also adopted 
timestamp based protocols to ensure the serializability.

\section{Conclusion}

In this paper, we proposed DGCC, a new dependency
graph based concurrency control protocol.
DGCC separates concurrency control from execution by building
dependency graphs for batches of transactions in order to resolve contention before execution.
We showed that DGCC can better exploit modern multicore hardware by having higher parallelism.
DGCC also removes the need of centralized
control components thereby giving better scalability.
A prototype DGCC-based OLTP system has been built that also
seamlessly integrated an efficient recovery mechanism.
Our extensive experimental study on YCSB and TPC-C shows that
DGCC achieves a throughput that is four times higher than
the classical concurrency control protocols for workloads with high contention.


{
\bibliographystyle{abbrv}
\bibliography{vldb_sample}  
}

\end{document}